\documentclass[onecolumn]{aa}
\usepackage{graphicx}
\usepackage{txfonts}
\begin{document}
   \title{uvby-$\beta$ photometry of high-velocity and metal-poor stars}

   \subtitle{XI. Ages of halo and old disk stars\thanks{Based on observations 
             collected at the H.L.~Johnson $1.5\,$m telescope at the Observatorio 
             Astron\'omico Nacional at San Pedro M\'artir, Baja California, 
             M\'exico.}{$^{,}$}\thanks{Tables 1--4 are only available in electronic
	     form at the CDS via anonymous ftp to cdsarc.u-strasbg.fr (130.79.128.5)
	     or via http://cdsweb.u-strasbg.fr/cgi-bin/qcat?J/A+A/(vol)/(page)}
	     }

   \author{W.J. Schuster\inst{1}, A. Moitinho\inst{2}, A. M\'arquez\inst{3},
          L. Parrao\inst{4}, \and  E. Covarrubias\inst{5}
          }

   \offprints{W.J. Schuster}

   \institute{Observatorio Astron\'omico Nacional, UNAM, Apartado Postal
              877, Ensenada, B.C., M\'exico, C.P. 22800\\
              \email{schuster@astrosen.unam.mx}
	 \and
	      CAAUL, Observat\'orio Astron\'omico de Lisboa, Tapada da Ajuda,
              1349-018 Lisbon, Portugal\\
	      \email{andre@oal.ul.pt}
	 \and
	      Instituto Nacional de Astrof\'{\i}sica, \'Optica y Electr\'onica, Luis
	      Enrique Erro No.~1, Tonantzintla, Puebla, M\'exico, C.P. 72840\\
	      \email{amarquez@inaoep.mx}
	 \and
	      Institute of Astronomy, UNAM, M\'exico, D.F., M\'exico, C.P. 04510\\
	      \email{laura@astroscu.unam.mx}
	 \and
	      Universidad de las Americas, San Pedro Cholula, Puebla, M\'exico,
	      C.P. 72820\\              
              }

   \date{Received July 08, 2005; accepted September 13, 2005}

   \abstract{New $uvby$--$\beta$ data are provided for 442 high-velocity and 
metal-poor stars; 90 of these stars have been observed previously by us, and 352 
are new. When combined with our previous two photometric catalogues, the data 
base is now made up of 1533 high-velocity and metal-poor stars, all with 
$uvby$--$\beta$ photometry and complete kinematic data, such as proper motions and 
radial velocities taken from the literature. Hipparcos, plus a new photometric 
calibration for M$_v$ also based on the Hipparcos parallaxes, provide distances 
for nearly all of these stars; our previous photometric calibrations give values 
for E(b-y) and [Fe/H].  The [Fe/H],V(rot) diagram allows us to separate these 
stars into different Galactic stellar population groups, such as old-thin-disk, 
thick-disk, and halo. The X histogram, where X is our stellar-population 
discriminator combining V(rot) and [Fe/H], and contour plots for the [Fe/H],V(rot)
diagram both indicate two probable components to the thick disk.  These population 
groups and Galactic components are studied in the (b-y)$_o$,M$_v$ diagram, 
compared to the isochrones of Bergbusch \& VandenBerg (2001), to derive stellar
ages. The two thick-disk groups have the mean characteristics: 
([Fe/H], V(rot), Age, $\sigma_{W'}$) $\approx$ 
(-0.7 dex, 120 km/s, 12.5 Gyr, 62.0 km/s), and 
$\approx$ (-0.4, 160, 10.0, 45.8). The seven most metal-poor halo groups, 
$-2.31 \leq$ [Fe/H] $\leq -1.31$, show a mean age of $13.0 \pm 0.2$ (mean error) 
Gyr, giving a mean difference from the WMAP results for the age of the Universe 
of $0.7 \pm 0.3$ Gyr. These results for the ages and components of the thick 
disk and for the age of the Galactic halo field stars are discussed in terms of 
various models and ideas for the formation of galaxies and their stellar 
populations.

   }
   \authorrunning{W. J. Schuster et al.}
   \titlerunning{Ages of halo and old disk stars}
   \maketitle
%
%________________________________________________________________

\section{Introduction}

In this day and age of Hubble deep fields, large-scale surveys (for example, Newberg
et al.~2002, Schneider et al.~2003, Yanny et al.~2003), WMAP (Bennett et al.~2003),
and $\Lambda$CDM models of hierarchical galaxy formation, astronomers and cosmologists
are moving closer to an understanding of the fundamentals of galaxy formation and
early evolution.  Within our Galaxy, the detection of the accreting dwarf galaxies
in Sagittarius (Ibata et al.~1994) and Canis Major (Martin et al.~2004, Bellazzini et
al.~2004) have provided observational evidence and details for accretion and merger
processes.  Tidal debris in the Galaxy has been detected and studied (Ibata et al.~2001,
2003; Newberg et al.~2002; Crane et al.~2003; Yanny et al.~2003; Rocha-Pinto et al.~2003,
2004; Majewski et al.~2004), as well as structure within the thick disk (Gilmore et 
al.~2002; Parker et al.~2003, 2004) and evidence of counterrotating kinematics of
thick disks in external galaxies (Yoachim \& Dalcanton 2005).  Models tying thick-disk
and halo characteristics to hierarchical clustering have been developed (Abadi et al.~2003,
Brook et al.~2004).  What is shown in this paper is that the data bases developed over
the last couple decades for the local high-velocity and metal-poor stars (Schuster
\& Nissen 1988, Schuster et al.~1993, Carney et al.~1994, Ryan 1989) have not yet been
totally exploited and exhausted for results concerning the above themes:  the age and
origins of the Galaxy and its halo; the age, structure, and formation of the Galactic
thick disk.  In this paper the $uvby$--$\beta$ photometry for more than 1500
high-velocity and metal-poor stars is applied to these problems of ages, structure,
and origins relating to our Galaxy and its stellar populations.

The $uvby$--$\beta$ photometric system is particularly suited for the study of 
high-velocity and metal-poor (hereafter, HV) F- and G-type stars, as has 
already been pointed out in several papers of this series, such as Schuster \&
Nissen (1988, SN; 1989a, Paper II; 1989b, Paper III), Nissen \& Schuster (1991,
Paper V), Schuster et al.~(1993, SPC), and M\'arquez \& Schuster (1994, MS).  Briefly,
intrinsic-color calibrations, $(b-y)_{\rm 0}$--$\beta$, exist that allow accurate 
and precise, $\pm 0\fm01$, measures of interstellar reddening excesses, $E(b-y)$, 
for individual field stars; such a calibration has been given in Paper II.  
In this same paper photometric calibrations are given for the [Fe/H]
values of F- and G-type stars.  Photometric absolute magnitudes and distances can
be calibrated and used effectively, as shown in Olsen (1984) and
Paper V.  This photometric system has the great advantage that
it permits us to obtain accurate stellar distances even for evolving main-sequence
and subgiant stars due to the gravity sensitivity of the $c_{\rm 0}$ index.  Also,
importantly, theoretical isochrones in the $T_{\rm eff}$,$M_{\rm bol}$diagram
can be transformed to the $(b-y)_{\rm 0}$,$M_{\rm V}$ or $(b-y)_{\rm 0}$,
$c_{\rm 0}$ diagrams for the estimation of relative and/or absolute ages
of evolving field stars that are near their respective turn-offs, and in 
several of the previous papers of this series the isochrones of VandenBerg and 
co-workers have been used for such purposes, to study the Galactic halo population and 
to make comparative analyses between the relative ages of the halo and thick-disk
stellar populations.  Most recently the isochrones of Bergbusch \& VandenBerg
(2001) have been transformed to the $uvby$ photometric system using the
color--$T_{\rm eff}$ relations of Clem et al.~(2004).  

$uvby$--$\beta$ photometry also can provide basic stellar
atmospheric parameters as a prelude to detailed chemical abundance studies
making use of high-resolution spectroscopy and model atmospheres.  Several
empirical calibrations already exist in the literature for the conversion
of $(b-y)_{\rm 0}$ or H$\beta$ to $T_{\rm eff}$ (for example, Alonso et al.~1996, 
1999; Ram\'irez \& Mel\'endez 2005a, 2005b); these calibrations include 
appropriate metallicity dependences.  Index diagrams, such as 
$(b-y)_{\rm 0}$,$c_{\rm 0}$, or the reddening-free $[m_{\rm 1}]$,$[c_{\rm 1}]$,
or $\beta$,$[c_{\rm 1}]$, allow the classification of field stars according
to their evolutionary status, permitting us to estimate the stellar surface 
gravities.  This information can also be used as input into the model-atmosphere 
analyses.  

In this paper, $uvby$--$\beta$ photometry is presented for 442 HV stars, 90 repeated 
from the previous SN and SPC catalogues and 352 observed for the first time by us;
in Sect.~2 the selection criteria of these stars, the observing and reduction
techniques for the $uvby$--$\beta$ data, comparisons of this data with our previous
photometric catalogues, and a description of this newest catalogue (Table 1) are
given.  In Sect.~3 are detailed our corrections for interstellar reddening, our
photometric abundances ([Fe/H]), the absolute magnitudes and distances, and our
photometric classifications (Table 2) for our total data set.  Section 4 describes 
the kinematic data and Galactic velocities, and Sect.~5, the cleaning of our total 
sample of binary and variable stars.  Section 6 concentrates on the [Fe/H],V(rot) 
diagram:  its appearance and use to separate stellar populations and groups, the 
stellar-population parameter, X, and its histogram, and contour plots and probable 
structure in the thick disk.  In Sect.~7, several $(b$--$y)_{\rm 0}$,$M_{\rm V}$ 
diagrams with isochrones are given for various thick-disk and halo groupings, plus the 
[Fe/H],$(b$--$y)_{\rm 0}$ diagram for all the halo stars of this data base together
with the turn-off, main-sequence, and subgiant stars of our very-metal-poor sample;
the bluer metal-poor stars and relative stellar ages are discussed.  Section 8 provides
a comparison of our ages in the form of an age versus [Fe/H] diagram, as well as
analyses of these ages, the thick-disk structure, and the bluer metal-poor stars in
terms of the WMAP results, thick-disk formation scenarios, and hierarchical clustering
and merging theories for galaxy formation in a $\Lambda$CDM universe.  Finally,
Sect.~9 sums it all together, giving our main conclusions.    

%__________________________________________________________________

\section{Photometric observations of the high-velocity and metal-poor stars}

\subsection{Selection of the stars}

The stars in the catalogue of this publication, Table 1, have been drawn from
several sources.  B.W. Carney sent a list of the bluer stars from their
survey (Carney et al.~1994) which lacked $uvby$--$\beta$ photometry, for the
purpose of determining stellar photometric ages.  S.G. Ryan sent a list of numerous
stars lacking $uvby$--$\beta$ included in various of their spectroscopic studies. 
The student F. Valera helped draw up a list of stars with [Fe/H] $\leq -2.00$ from
the literature, including the SN and SPC catalogues, the Carney et al. (1994)
survey mentioned above, the lists of Sandage \& Fouts (1987), as well as those 
of Norris \& Ryan (1989), the idea being to obtain five or more independent 
$uvby$--$\beta$ observations for each of these more metal-poor stars to provide 
more precise stellar atmospheric parameters for various spectroscopic projects.  
Stars lacking $uvby$--$\beta$ photometry were also selected from the lists of 
Thorburn (1994).

\subsection{Observation and reduction techniques}

The $uvby$--$\beta$ data presented here in Table 1 for the HV stars were taken
using the H.L. Johnson $1.5\,$m telescope at the San Pedro M\'artir Observatory,
Baja California, M\'exico (hereafter SPM), and the same six-channel $uvby$--$\beta$ 
photoelectric photometer as for the northern observations of SN, all the
$uvby$--$\beta$ observations of SPC, the northern data of very-metal-poor stars
by Schuster et al.~(1996), and the $uvby$--$\beta$ data for very-metal-poor stars
in Table 1 of Schuster et al.~(2004, Paper X).  The new $uvby$--$\beta$ values 
included in Table 1 were taken during ten observing runs from April 1991 through
November 1997.

The $uvby$--$\beta$ data presented here for the HV stars in Table 1 were taken 
and reduced using techniques very nearly the same as for SN and SPC;
see these previous papers for more details. The four-channel $uvby$ section 
of the SPM photometer is really a spectrograph-photometer that employs exit 
slots and optical interference filters to define the bandpasses.  The grating
angle of this spectrograph-photometer was adjusted at the beginning of each 
observing run to position the spectra on the exit slots to within about 
$\pm 1{\AA}$.  Whenever possible, extinction-star observations were made
nightly over an air-mass interval of at least 0.8 (see Schuster \& Parrao 2001;
also Schuster et al.~2002),
and spaced throughout each night several ``drift'' stars were observed
symmetrically with respect to the local meridian.  Using these observations
the atmospheric extinction coefficients and time dependences of the night
corrections could be obtained for each of the nights of observation
(see Gr{\o}nbech et al.~1976).  Finding charts were employed at SPM whenever
available .  As for previous studies, such as SN and SPC, the program stars
were observed at SPM to at least 50,000 counts in all four channels of $uvby$
and to at least 30,000 counts for the two channels of H$\beta$.  For all
program stars the sky was measured until its contributing error was
equal to or less than the error of the stellar count.  At SPM an attempt
was made to obtain three or more independent $uvby$ observations for each
of the program stars, and at least five independent observations for
those with [Fe/H] $\leq -2.00$.

As for the SN and SPC catalogues, all of these data reductions were carried
out following the precepts of Gr{\o}nbech et al.~(1976) using computer 
programs kindly loaned by T. Andersen (see Parrao et al.~1988).  At SPM the
$uvby$--$\beta$ standard stars observed were taken from the same lists as
for the previous catalogues; these are mostly secondary standards from
the catalogues of Olsen (1983,1984).  The reduction programs
create a single instrumental photometric system for each observing run,
including nightly atmospheric extinctions and night corrections with
linear time dependences.  Then transformation equations from the
instrumental to the standard systems of $V$, $(b$--$y)$, $m_{\rm 1}$, $c_{\rm 1}$,
and $\beta$ are obtained using all standard stars observed during that
observing period.  The equations for the transformation to the standard
$uvby$--$\beta$ system are the linear ones of Crawford \& Barnes (1970)
and of Crawford \& Mander (1966).  Small linear terms in $(b-y)$
are included in the standard transformation equations for $m_{\rm 1}$
and $c_{\rm 1}$ to correct for bandwidth effects in the $v$
filter.  Our $y$ measures were transformed onto the V system of Johnson
et al.~(1966).

%                                                One column figure
%------------------------------------------------------CompPhotVsV
   \begin{figure}
   \centering
   \includegraphics[width=12cm]{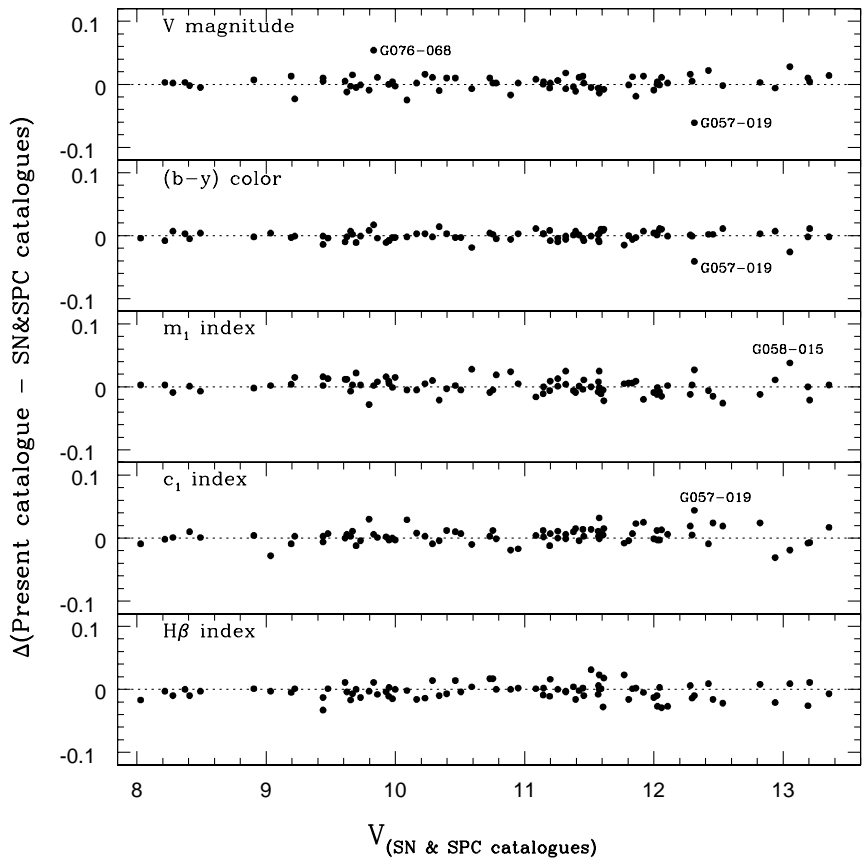}
      \caption{Comparison of the new $uvby$--$\beta$ data, taken for
      the present catalogue, with previous values published in the
      SN and SPC catalogues, for those 90 stars which have been
      repeated.  The comparison is shown in the sense:  the difference
      between the newer values minus the older values of SN and/or SPC as a 
      function of the previous $V$ magnitudes.  The dotted lines show the 
      zero-difference levels, and for three of the stars which are possible 
      photometric variables ($\Delta \geq 0\fm04$), their points have been 
      labeled with their identification numbers.
              }
         \label{FigCompPhotVsV}
   \end{figure}
%
%______________________________________________________________
%                                                One column figure
%------------------------------------------------------CompPhotVsBY
   \begin{figure}
   \centering
   \includegraphics[width=12cm]{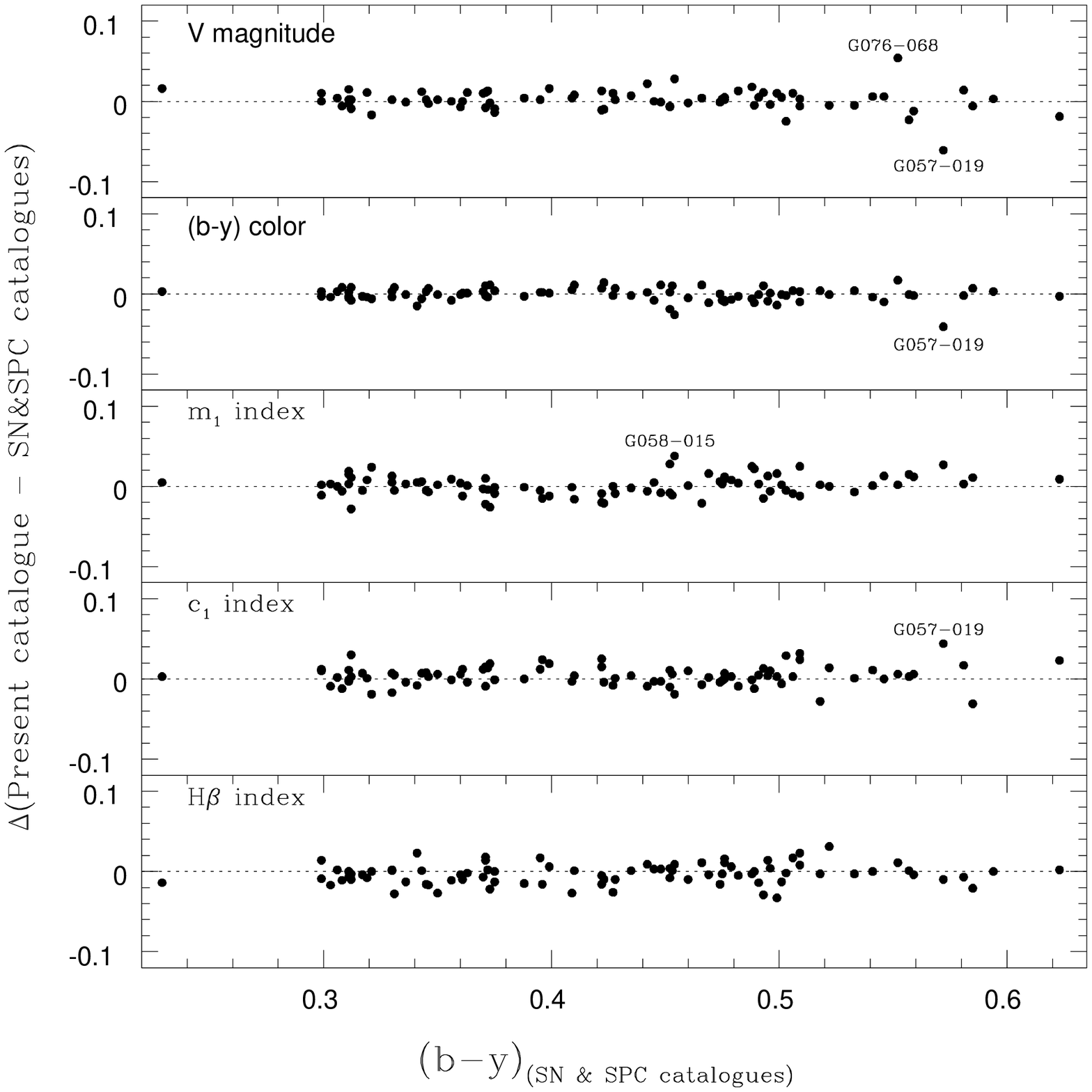}
      \caption{Same as the previous figure, but as a function of the
      $(b$--$y)$ values of SN and/or SPC.
              }
         \label{FigCompPhotVsBY}
   \end{figure}
%
%______________________________________________________________

For the SPM $uvby$--$\beta$ data of Table 1, typical (median) values 
for the standard deviations of a single observation are $\pm0\fm008$, $0\fm005$--$6$,
$0\fm007$--$8$, $0\fm008$--$9$ and $0\fm008$ for $V$, $(b$--$y)$, $m_{\rm 1}$, $c_{\rm 1}$
and $\beta$, respectively.

\subsection{The catalogues of observations}

Table 1 presents the $uvby$--$\beta$ catalogue for the 442 HV stars observed
at SPM.   Column 1 lists the stellar identifications according to several
different nomenclatures, mainly DM numbers, NLTT (Luyten 1979a), and the Lowell 
proper-motion survey (Giclas et al.~1959, 1961, ..., 1975); Col.~2, the $V$
magnitude on the standard Johnson $UBV$ system; and Cols.~3--5 and 7, the 
indices $(b$--$y)$, $m_{\rm 1}$, $c_{\rm 1}$ and $\beta$ on the
standard systems of Olsen (1983, 1984), which are essentially the systems
of Crawford \& Barnes (1970) and Crawford \& Mander (1966) but with a
careful extension to metal-poor stars and with north-south systematic
differences corrected.  Columns 6 and 8 give $N_u$ and $N_{\beta}$, the
total numbers of independent $uvby$ and $\beta$ observations.
Stars marked with an asterisk, ``*'', in the last column have
observing notes and cross-identifications in the ``Notes'' section
following the main table, as described in the next paragraphs.

In the second section of Table 1 notes for the HV stars, taken during the
observations or during the data reduction and analysis, are given; also listed
are many cross-identifications, especially from the HD, DM, LTT (Luyten 1957,
1961), LHS (Luyten 1979b), and
Lowell proper-motion catalogues.  Many of the observing notes concern the
photometric contamination, or possible contamination, by a fainter nearby
star, as indicated by a note ``D'' indicating duplicity.  For example, 
LP808-022 was observed in a very crowded field with
a 20 arcsecond diaphragm; probably its ``star'' and ``sky'' measures were
contaminated by fainter nearby stars, and the note ``D" plus the accompanying
text indicate this problem.  LP811-024 and LP866-018 were also observed in
crowded fields, as their texts indicate.  For LP920-024 the ``D" and text 
show that it was observed with a fainter star located just outside the 
diaphragm; the contamination is probably very small, due to scattered light 
from this fainter star.  The star G006-013 was observed slightly off-center 
in the diaphragm to place a faint star just outside the diaphragm; since the 
bandpasses of the SPM photometer are mainly filter-defined, this small offset 
should produce negligible errors for the indices.  G096-001 was observed with
a second star within the diaphragm; in this case the photometric contamination
is clear.

The notes ``SN'' and ``SPC'' mark those stars which have been repeated from
these two previous $uvby$--$\beta$ catalogues; these are the stars plotted in
Figs.~1 and 2, and the values of Table 1 combine the newer $uvby$--$\beta$
observations with the older.  Four stars (+39:2173 = G115--034, +34:0796 = GC4849,
+26:2621 = G166--054, and $-$15:2546 = HD74000) were sent to us by different
astronomers to obtain $uvby$--$\beta$ data using these different identification 
numbers, and were observed by us as separate stars until the cross-identifications
were checked; \emph {all} the independent observations for each of these
four stars have been combined for the values of Table 1, but these stars 
appear twice in Table 1 using both of the original identifications
sent to us, plus the note ``repeated''.  The note ``ID?'' appears for those
stars identified only from their coordinates, without any finding chart.  The
text ``Need confirming observations'' marks stars observed only once for the
values of Table 1.  For these last two categories, ``ID?'' and ``Need ...'',
the values of Table 1 generally show clear evidence of low photometric
metallicities, i.e. low $m_{\rm 1}$ values for the corresponding $(b$--$y)$, 
indicating that the correct HV stars have indeed been observed.

Stars with ``++ Red subgiant/giant'', such as LP685--047 and G013--050, have
$(b$--$y) \geq 0\fm50$ and $c_{\rm 1} \geq 0\fm35$; as discussed in SN,
the $m_{\rm 1}$ and $c_{\rm 1}$ values of these stars may be less accurate,
due to non-linearity of the photometric transformations, and the  $m_{\rm 1}$ 
and $c_{\rm 1}$ values in Table 1 have been placed in parentheses for such
stars.  Stars with $(b$--$y)$ and $c_{\rm 1}$ values close to these limits,
but not both exceeding, have a note of ``+'', such as LP920--024 and G012--022.

For the cross-identifications, some of the stars from the Lowell-proper-motion
survey (the ``G'' stars) have been identified in several fields, as many as
five in some cases.  We give a maximum of two ``G'' identifiers for each star,
and if a given star has more, the second is followed by an ellipsis (``G166--054...'',
for example).

\subsection{Comparisons with our previous photometric catalogues}

For 90 of the stars in Table 1, new $uvby$--$\beta$ observations have been
taken for the present catalogue, as well as their older observations presented
in the SN and SPC catalogues.  These stars are marked in the ``Notes''
section of Table 1 with ``SN'' or ``SPC''.  Comparisons of this newer and
older $uvby$--$\beta$ data are shown in Figs.~1 and 2:  in Fig.~1 as a function
of the older $V$ magnitudes, and in Fig.~2 as a function of the older $(b$--$y)$
colors.  In these figures it is seen that the quality of the newer $uvby$--$\beta$
data is quite good and closely on the same photometric systems as the older
catalogues of SN and SPC.  Only a few of the new observations have residuals
greater than $0\fm04$ compared to the older values, and these cases have been
labeled with the stars' names in Figs.~1--2, as candidates for variable stars.
The label ``G057--019'' appears thrice, and ``G058--015'' and ``G076--068''
once each.

The average difference for the comparison of the $V$ magnitudes in Figs.~1--2
is $+0.0024$ with a scatter of $\pm0.0116$; for the $(b$--$y)$ color,
$-0.0005 \pm0.0074$; for the $m_{\rm 1}$ index, $+0.0010 \pm0.0124$; for $c_{\rm 1}$,
$+0.0034 \pm0.0118$; and for H$\beta$, $-0.0033 \pm0.0126$.

%                                                One column figure
%-----------------------------------------------------------FeHhisto
   \begin{figure}
   \centering
   \includegraphics[width=12cm]{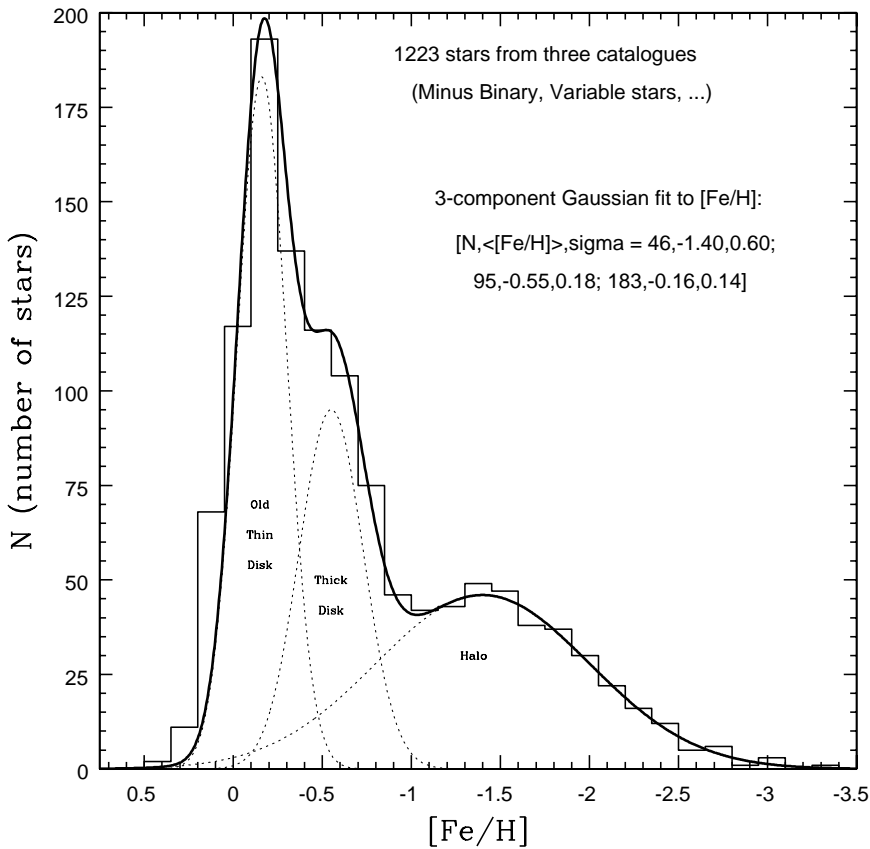}
      \caption{The [Fe/H] histogram for our total sample of
      1223 stars, cleaned of binary and variable stars as
      described in the text. The [Fe/H] values have been derived
      from the calibration of Paper II, and the
      bin size is 0.15 dex.  A three-component Gaussian fit to
      this histogram has been made using mean values and dispersions
      from the literature for the halo, thick disk, and old thin
      disk, such as those values from SPC.
              }
         \label{FigFeHhisto}
   \end{figure}
%
%______________________________________________________________
%                                                One column figure
%------------------------------------------------------CompMvsPhotHipp
   \begin{figure}
   \centering
   \includegraphics[width=12cm]{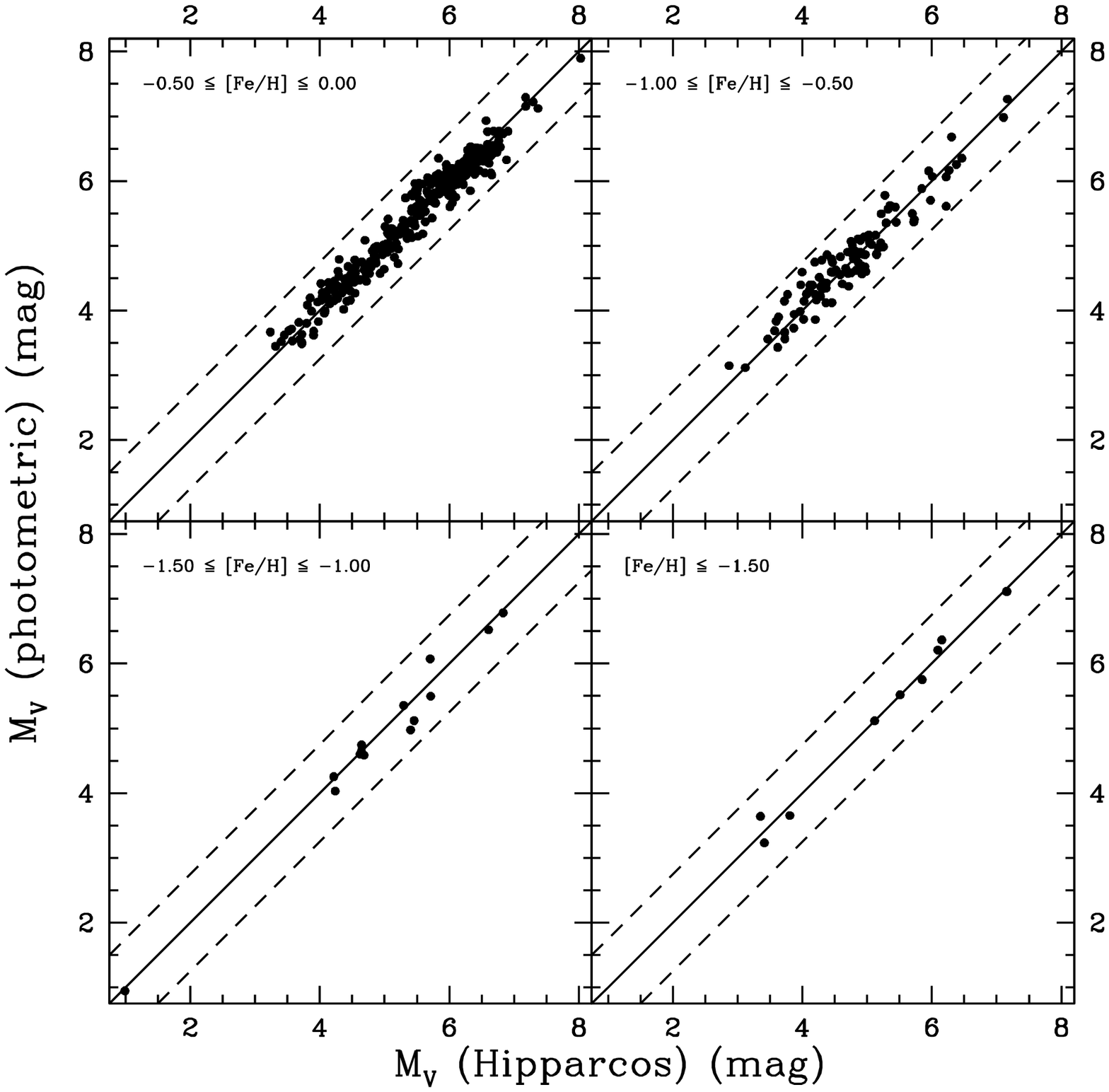}
      \caption{$M_{\rm V}$ values (photometric and Hipparcos) are 
      compared for four metallicity intervals:  $M_{\rm V}$ calculated 
      using $uvby$ photometry plus our empirical, Hipparcos-based, 
      photometric calibration are plotted against $M_{\rm V}$ taken 
      directly from the Hipparcos parallaxes.  The solid, diagonal 
      lines show one-to-one relations, and the dashed lines show 
      displacements of three-quarters of a magnitude.  The [Fe/H]
      values have been taken from the calibration of Paper II.
              }
         \label{FigCompMvs}
   \end{figure}
%
%______________________________________________________________

\section{Reddening, abundances, distances, and classifications}

\subsection{Interstellar reddening}

These stars can be dereddened using the intrinsic-color calibration of
Paper II when a value has been observed for H$\beta$, as for
most of the HV stars.  This calibration, plus a small offset correction
as noted by Nissen (1994)\footnote{ Nissen compared Na I column densities
from interstellar Na I lines with the color excesses derived from
Str\"omgren photometry plus the intrinsic-color calibration.  For 23 stars
without detectable interstellar Na I lines the mean value of $E(b$--$y)$
was slightly positive, +0.005, suggesting that the zero-point in the
$(b$--$y)_{\rm 0}$-$\beta$ calibration of Paper II should be increased by
0.005 mag.},
has been used to estimate interstellar reddenings
for the 442 high-velocity and metal-poor stars studied here (and also those
stars from the SN and SPC catalogues).  In Paper X
(Fig.~3) a comparison of the interstellar reddenings from this calibration
are compared to values derived from the maps of Schlegel et al.~(1998).  This
comparison is quite satisfactory with no evidence for systematic differences.
In this paper, reddening corrections from our intrinsic-color calibration have
been applied to the $uvby$ photometry only when
$E(b$--$y) \geq 0\fm015$; values smaller than this are mostly not real but are 
due to the photometric observational errors (see Nissen 1994).  For the other 
reddening corrections, these relations have been used:  $A_{\rm V} = 4.3E(b$--$y)$, 
$E(m_{\rm 1}) = -0.3E(b$--$y)$ and $E(c_{\rm 1}) = +0.2E(b$--$y)$ 
(Str\"omgren 1966; Crawford 1975).

\subsection{Photometric stellar abundances}

The metallicities, [Fe/H] values, used in this study have been derived from
the empirical calibrations of Paper II, one for F-stars and one for G-stars.  
The reader is referred to this earlier paper for more details.  For these
empirical calibrations the estimated standard deviations of a single 
photometric determination of [Fe/H] are $\pm 0.14$ at [Fe/H] $\approx -0.5$, 
$\pm 0.21$ at [Fe/H] $\approx -1.5$, and $\pm 0.30$ at [Fe/H] $\approx -2.5$.  
Comparisons by Feltzing et al.~(2001) of photometric abundances using our 
calibration equations with abundances from two recent spectroscopic studies 
have shown that these error estimates are overly conservative; they find a 
scatter of $\pm 0.10$--0.11 for the more metal-rich group.  Twarog et al.~(2002) 
suggest a systematic error in our G-star [Fe/H] calibration for 
[Fe/H] $\ga$ -0.2 and $(b$--$y) \ga 0.47$, but since the present paper concerns 
only the ages of ``turn-off" thick-disk and halo stars (i.e. the bluer stars) 
with [Fe/H] $<$ -0.2, this possible systematic difference contributes no problem 
to the following analyses and conclusions.  Also, the comparisons of Feltzing et 
al.~(2001) for stars with $-1.0 \la$ [Fe/H] $\la +0.3$ show systematic 
differences of our photometric [Fe/H] values of only +0.01--0.02, when compared 
to the recent spectroscopic studies in the sense 
[Fe/H]$_{\rm Phot} -$ [Fe/H]$_{\rm Sp}$.  And, the comparisons of Martell \& 
Smith (2004; Martell \& Laughlin 2002) with the results from their even more 
recent photometric calibrations for [Fe/H] suggest systematic problems with our
older calibration equations are smaller (in the absolute value) than 0.027,
for stars with $-0.4 \la$ [Fe/H] $\la +0.4$.  Figure 4 of Karata\c{s} et al.~(2005)
also shows very good agreement between our photometric [Fe/H] values and recent
spectroscopic ones over the range $-2.0 \la$ [Fe/H] $\la -0.2$.

In Fig.~3 is shown the [Fe/H] histogram for the 1223 high-velocity and
metal-poor stars of our total sample, from the SN, SPC, and present catalogues,
cleaned of the binary and variable stars as described below.  This plot is very
similar to Fig.~1b of SPC.  The [Fe/H] histogram is fit very well with three
Gaussian components corresponding to the old thin disk, thick disk and halo,
with the mean values and widths selected from the literature, especially SPC
and Mihalas \& Binney (1981).  The values used in Fig.~3 are
($\langle$[Fe/H]$\rangle$,$\sigma_{\rm [Fe/H]}$) = ($-1.40$,0.60) for the halo
Gaussian, ($-0.55$,0.18) for the thick disk, and ($-0.16$,0.14) for the old thin
disk.  As noted in SPC, the third component, the thick disk, is very much
needed for a good Gaussian fit to this [Fe/H] histogram.

\subsection{Absolute magnitudes and distances}

The absolute magnitudes and distances for this study are calculated directly 
from the Hipparcos (ESA 1997) parallaxes when these have errors of 10\% or less.
When these errors are larger or the stars not found in Hipparcos, the $M_{\rm V}$
and photometric distances are derived from an empirical calibration based upon
Hipparcos data (ESA 1997), described below and previously in Paper X.  For stars 
lying outside the range of this empirical calibration, Hipparcos parallaxes with 
errors larger than 10\% or absolute magnitudes from the older calibration of
Paper V are also used to obtain distances.  In the analyses to follow, of the
1533 stars with complete kinematic data, only 24 have had their distances 
derived from these last two, less exact, sources:  7 from Hipparcos with errors
greater than 10\% and 17 from the older calibration of Paper V.

Our empirical calibration equation for $M_{\rm V}$ is based upon 512 stars
from the Hipparcos data base with parallax errors of 10\% or less.  The 
Lutz-Kelker corrections to $M_{\rm V}$ for these stars are less than about 0\fm12 
(Lutz \& Kelker 1973).  This sample has been cleaned of binaries using other data 
bases (principally Dommanget \& Nys (1994) from SIMBAD; see Sect.~5 for further
details on the cleansing of binaries) and also by an iterative procedure
whereby stars with residuals $\ga 0\fm7$ in the calibration comparison have been
removed.  The calibration equation is a polynomial in $(b$--$y)$, $c_{\rm 0}$
and $m_{\rm 0}$ and higher-order terms to the fourth order.  As for the
calibrations of Paper II, the solution has been iterated until
all terms have T-ratios with absolute values greater than 3.  That is, all
coefficients are at least three times their estimated errors according to the
IDL REGRESS routine (the returned errors of the coefficients are standard
deviations). The data have $\sim 500$ degrees of freedom and so all coefficients
are non-zero at a significance level greater than 0.995.  The final regression
equation shows a scatter of $\pm 0\fm206$ in $M_{\rm V}$.  The 512 calibration
stars have the following ranges:  -2.4 $\la$ [Fe/H] $\la$ +0.4,
$0\fm038 \leq m_{\rm 0} \leq 0\fm593$, $0\fm279 \leq (b$--$y) \leq 0\fm600$,
$0\fm102 \leq c_{\rm 0} \leq 0\fm474$ and $0\fm991 \leq M_{\rm V} \leq 8\fm029$.
The actual region in the $(b$--$y)$,$M_{\rm V}$ diagram over which this
calibration is well-defined is a somewhat irregular polygon and not a rectangle;
it is good for main-sequence, evolving-main-sequence, and turn-off stars over
the range in $(b$--$y)$ given above, and for subgiant stars to
$(b$--$y)$,$M_{\rm V} \approx (0.42,+0.5)$ and $\approx (0.52,+4.0)$.
As mentioned in previous papers, for the most metal-poor stars many
photometric calibrations are not entirely adequate since few good calibration
stars with [Fe/H] $< -2.0$ exist.  This caveat also applies here, but this
Hipparcos-based photometric calibration seems to work quite well for the very
metal-poor stars as shown in Paper X of this series.  It certainly works very well
to [Fe/H] $\approx -2.40$, which is the approximate metal-poor limit of the stars 
analyzed in this paper.

In Fig.~4, $M_{\rm V}$ values calculated using our Hipparcos-based, empirical
calibration are plotted against $M_{\rm V}$ derived directly 
from the Hipparcos parallaxes for four metallicity intervals:  $-0.50 \leq$ [Fe/H]
$\leq 0.00$, $-1.00 \leq$ [Fe/H] $\leq -0.50$, $-1.50 \leq$ [Fe/H] $\leq -1.00$,
and [Fe/H] $\leq -1.50$, according to the photometric [Fe/H] calibration of 
Paper II.  The nine stars of the last panel have [Fe/H] 
values in the range $-2.39 \leq$ [Fe/H] $\leq -1.57$, are the same stars plotted
in Fig.~7 of Paper X, and show that our calibration stars do not extend to the
lowest [Fe/H] values of some of the HV stars, but do cover well the metallicities 
of the isochrones used in this publication.  The agreement seen in Fig.~4 is quite
satisfactory for all four panels.

In the analyses to follow, three different symbols are used to denote the $M_{\rm V}$s
and distances with different sources and qualities.  Stars with values direct from
Hipparcos with errors less than 10\% are shown as circles, those with values from the
empirical, IDL-Hipparcos photometric calibration, as squares, and those from the older
calibration of Paper V or from Hipparcos but with larger errors, as triangles.

\subsection{The $(b$--$y)_{\rm 0}$,$c_{\rm 0}$ diagram and photometric
            classifications}

In Fig.~6 of Paper X a $(b$--$y)_{\rm 0}$,$c_{\rm 0}$ diagram is presented for
very-metal-poor (VMP) stars, and a scheme is suggested for classifying these VMP
stars into categories such as those identified in globular cluster CM diagrams:  turnoff
stars (TO), main sequence stars (MS), blue stragglers (BS), the transition from
blue-straggler to turnoff (BS-TO), subgiant stars (SG), red giants (RG), horizontal
branch stars (HB), the transition red-horizontal-branch to asymptotic-giant-branch
(RHB-AGB), blue horizontal branch (BHB), subluminous stars (SL), and the transition 
from the blue-horizontal-branch to subluminous (SL-BHB).  This scheme is also useful
for classifying  the evolutionary states of stars from the present paper, as well as
stars from the SN and SPC catalogues.
However, the stars in Fig.~6 of Paper X have a mean metallicity of
[Fe/H] $\approx -2.4$, and the stars studied here range from above-solar [Fe/H] 
values to [Fe/H] $\approx -5.0$.  To adjust for a possible systematic error in the 
photometric classifications, the $c_{\rm 0}$ values of the stars have been corrected
according to their [Fe/H] and $(b$--$y)_{\rm 0}$ values using Eq.~2 of Paper V to 
obtain $c_{\rm 0}$ values corresponding to [Fe/H] $= -2.40$.  According to Paper V, this
Eq.~2 is valid only over the range $+0.27 \la (b$--$y)_{\rm 0} \la +0.58$.  The
corrections applied to $c_{\rm 0}$ are small; the median of the absolute values is
$\approx 0.050$, and the mean $\approx 0.064$, excluding the two stars in Table 2 with
a Note of ``B", indicating a very uncertain value for [Fe/H].  (A very
small shift in the $(b$--$y)_{\rm 0}$ values, due to the metallicity differences, 
has been ignored here.)  Also, according to Paper II, the stars' photometric [Fe/H]
values have been derived only over the range $+0.22 \la (b$--$y)_{\rm 0} \la +0.60$, 
and for stars not too evolved:  $+0.17 \la c_{\rm 0} \la +0.60$ for the F-stars and
$+0.10 \la c_{\rm 0} \leq +0.50$ for the G-stars.  Stars outside these ranges have
not had their $c_{\rm 0}$ values corrected to [Fe/H] $= -2.40$ prior to the
photometric classification.  Also from Paper II, the intrinsic-color calibration is 
useful only over the ranges, $+0.25 \la (b$--$y)_{\rm 0} \la +0.55$ and
$+0.12 \la c_{\rm 0} \la +0.54$; for stars outside these ranges,
$c_{\rm 0}$ and $(b$--$y)_{\rm 0}$ have been set equal to their observed values,
$c_{\rm 1}$ and $(b$--$y)$.

According to these corrections, color limits, and Fig.~6 of Paper X, most of the
stars in SN, SPC, and this paper classify as main sequence (MS) and turnoff (TO)
stars.  (In Fig.~6 of Paper X the line separating the MS and SL stars was very 
roughly drawn employing only a small number of stars, $\approx 10$; for the present
work a much larger sample of stars has shown the need to shift this line downward by 
about $0\fm00$ at $(b$--$y)_{\rm 0} = 0.30$, $0\fm10$ at $(b$--$y)_{\rm 0} = 0.40$, 
and $0\fm06$ at $(b$--$y)_{\rm 0} = 0.50$; with this shift the distinction between MS
and SL is clear.)  In Table 2 are given those stars \emph {not} classified as MS or TO:
in Column 1 is shown the stars' identification numbers; Cols.~2 and 3, the
$(b$--$y)_{\rm 0}$ and $c_{\rm 0}$ values, respectively; Col.~4, the photometric [Fe/H] 
value; Col.~5, the photometric classification from Fig.~6 of Paper X; and Col.~6, notes.
In this latter ``Notes" column, ``A" emphasizes stars without photometrically determined
[Fe/H] values; ``B" marks two stars with [Fe/H] $\approx -5.0$ (at such low values the
$m_{\rm 1}$ index is no longer sensitive to the metallicity, and so these [Fe/H] values
are very uncertain); ``C" shows two red giant stars with $(b$--$y) \ge +0.50$ and
$c_{\rm 1} \ge +0.35$ (as discussed in SN, such stars will have less accurate $c_{\rm 1}$
values due to transformation errors); and ``D" indicates stars which are repeated in this
table (4306 = HD4306, 25532 = HD25532, G014--032 = G014--032, W5793 = $-$12:2669, and
W6296 = +44:1910).

These stars of Table 2 represent the redder, bluer, and/or more evolved stars of our total
sample and are interesting for future studies, such as spectroscopic ones.  Many of
these stars fall outside the photometric calibrations and their color-limits, as 
discussed above and as can be seen in the fourth column, where many photometric [Fe/H] 
values are missing.  So in general the classifications are rough, but do serve to point 
out the more extreme, potentially interesting stars of our sample.  For example, six
stars have been classified ``SL'' (HD26, G038--001, HD108754, G015--024, G152--035, and
833--020), are photometrically unusual, and are candidates for being subluminous, i.e.
degenerate stars.  Twelve stars are candidate subgiant stars:  HD17072, $-$24:1782,
W6296 = +44:1910, HD103459, HD128204, HD140283, HD165271, HD200654, G092--006, G026--012,
811--024, and G170--047.  Another 24 stars are potentially even more evolved with eight
red-giant classifications, nine horizontal-branch candidates, four RHB-AGBs (HD25532,
HD117327, HD229274, and HD195636), and three BHBs (HD139961, HD213468, and HD214539).
Nine of the bluer stars are possible blue stragglers with ``BS'' or ``BS-TO'' classifications.

\section{Kinematic data and Galactic velocities}

\subsection{Proper motions and radial velocities}

The proper motions and coordinates used in this paper have been taken primarily from 
three sources:  Hipparcos (ESA 1997), Tycho--2 (H{\o}g et al.~2000), and the revised NLTT
(Salim \& Gould 2003).  For a few stars other sources have been employed, such as the 
original NLTT (Luyten 1979a), the LHS (Luyten 1979b), the Lowell Proper Motion Survey 
(Giclas et al.~1959, 1961, ..., 1975), and the PPM Star Catalogue (R\"oser \& Bastian 1991).  
Of the 1706 stars from our combined three catalogues of photometric data (SN, SPC, and 
this paper) proper motions have been found for all of the stars except three.  In general
the proper motion errors range from 0.5--3.0~mas/yr per component for the best sources,
such as Hipparcos and Tycho--2, to as large as $\ga 10$~mas/yr per component for the
other sources, such as the NLTT and the Lowell Proper Motion Survey.

The radial velocities for the present study have been obtained from the literature,
from a number of sources, such as Carney et al.~(1994), Barbier-Brossat et al.~(1994, 2000),
Nordstr\"om (2000), Ryan \& Norris (1991), Fouts \& Sandage (1986), and Abt \& Biggs (1972).
Of the stars from our three photometric catalogues radial velocities have been found for
all except 60 stars.  The observational errors of these velocities range from a few tenths
of a km s$^{-1}$ for the best sources, such as Carney et al.~(1994) and Nordstr\"om (2000),
to $\approx 7$ km s$^{-1}$ for the older sources.

\subsection{Galactic velocities}

There are 1706 stars in our combined three photometric catalogues, of which 90 have been
repeated and another 87 are without complete kinematic data, lacking either radial
velocities, distances, or proper motions, leaving 1533 stars with complete data for 
calculating the Galactic velocities, (U,V,W).  These velocities and their errors have 
been derived using the formulae and matrix equations from Johnson \& Soderblom (1987).
For these calculations, the celestial coordinates, proper motions, and radial velocities,
as well as the errors of the proper motions and radial velocities, have been taken from
the observational sources and compilations mentioned in the previous subsection.

The stellar distances have been derived from the absolute magnitudes as described above
in subsection 3.3:  from Hipparcos parallaxes (ESA 1997), from our present photometric
$M_{\rm V}$ calibration, which also depends upon Hipparcos, or from the older photometric
$M_{\rm V}$ calibration of Paper V.  The distance errors have been taken from the Hipparcos
parallax errors, when appropriate, or, for the newer empirical photometric $M_{\rm V}$
calibration, set equal to 10\% as a conservative estimate; the final regression equation
gives a scatter of $\pm 0\fm206$ for this $M_{\rm V}$ calibration.  For 17 stars with distances
from the older calibration of Paper V, the distance errors have been adjusted according 
to the evolutionary corrections, the $\delta M_{\rm V}$ of this previous paper.  
For example, for stars which are little evolved, $-0\fm25 < \delta M_{\rm V} < +0\fm25$, 
the distance error has been set at 10\%, while for stars very evolved, 
$\delta M_{\rm V} > +4\fm0$, at 30\%, with other levels of error in between.  This 
scaling of the distance error with evolutionary status of the star takes into account 
various systematic problems with this older $M_{\rm V}$ calibration.

The Galactic velocities from Johnson \& Soderblom (1987) correspond to a right-handed 
coordinate system with (U,V,W) positive in the directions of the Galactic center, Galactic
rotation, and the North Galactic Pole, respectively.  In the present paper 
(+10.0, +14.9, +7.7) km s$^{-1}$ have been used to correct (U,V,W) for the solar motion,
(U',V',W') = (U,V,W) + (+10.0, +14.9, +7.7), and our typical (median) errors in these 
velocities are $(\pm 6.2, \pm 6.6, \pm 4.8)$ km s$^{-1}$.
 
Tables 3 and 4 show the input and output values for these kinematic calculations,
respectively, plus Table 4 also gives other parameters used in the following analyses,
such as the dereddened photometry as described in Sect.~3.1, the photometric metallicities
from Sect.~3.2, the selected absolute magnitudes from Sect.~3.3, and the X parameter from
Sect.~6.1 below.  In Table 3 the first two columns give the stars' identifications, where
Col.~1 contains our internal numbers and Col.~2 identifications from various external
catalogues.  In Col.~1, numbers prefixed by ``SN'' refer to the $uvby$--$\beta$ catalogue 
of SN, prefixed by ``SPC'' the catalogue of SPC, and prefixed by ``SPG'' the present 
catalogue of Table 1; in each of these sections of Table 3 the stars are listed in the same
order, and in Col.~2 the preferred external identifications are the same, as in those 
original catalogues:  SN, SPC, and Table 1.  In Table 3, Cols.~3--5 and 6--8 give the right 
ascension and declination, Col.~9 the radial velocity in km s$^{-1}$, Cols.~10 and 11 the 
proper motions in right ascension and declination ($\mu_{\alpha}$ and $\mu_{\delta}$) in 
mas yr$^{-1}$, Cols.~12--14 the standard errors of the radial velocity, of $\mu_{\alpha}$, 
and of $\mu_{\delta}$ in the same units as above, respectively, Cols.~15--16 the distance in 
parsecs as described in Sect.~3.3 plus its estimated standard error, and finally in 
Col.~17 the epoch of these values.  

In Table 4 the fist two columns are the same as in Table 3, Cols.~3--5 contain
the Galactic velocities (U',V',W') in km s$^{-1}$, Cols.~6--8 the standard errors of these velocities, respectively, Cols.~9--10 once again the distance in parsecs plus its 
estimated standard error, Cols.~11--14 the dereddened photometry ($V_{\rm 0}, 
(b$--$y)_{\rm 0}, m_{\rm 0}, c_{\rm 0}$), Col.~15 the photometric metallicity ([Fe/H], 
where ``9.99'' indicates that no value has been obtained due to the limits of the 
calibrations given in Paper II), Col.~16 the selected absolute magnitudes from Sect.~3.3,
Col.~17 a code for these absolute magnitudes (where ``1'' indicates an absolute magnitude
derived directly from a Hipparcos parallax with a less than 10\% error, ``2'' an absolute
magnitude from the empirical calibration described in Sect.~3.3, and ``3'' those 
$M_{\rm V}$s from the older calibration of Paper V or from Hipparcos but with larger 
relative errors), Col.~18 the values of the X parameter from Sect.~6.1 (where 
``$-$999.999'' indicates no value since [Fe/H] has not been obtained), and Col.~19 a code 
for the binary and variable stars, as described in the following section (where ``1'' 
indicates a star not identified as a binary, ``2'' a double-lined spectroscopic binary, 
SB2 or dlsb, ``3'' all other binaries, suspected binaries, or those stars noted at the 
telescope to be contaminated, and ``4'' the probable photometric variables.  Table 4 
contains 1533 lines, corresponding to 1533 different stars, less than Table 3, since all
repetitions, between or within the different photometric catalogues, have been eliminated.

\section{Binary and variable stars}

Of the 1533 stars in our kinematic sample, many are known or suspected binary stars, and
a few are also known or suspected variable stars.  In addition, during the observations
of the $uvby$--$\beta$ photometry it was noted at the telescope that nearby fainter stars 
produced slight, some, or significant luminous contamination to the data being either just 
outside the photometer's diaphragm, at the diaphragm's edge, or well within, respectively.
These are noted as such in the second half of Table 1 of this paper;  the SN and SPC
catalogues also have similar ``Note'' sections to their tables.
To correct our sample for the possible errors and systematic effects that such contamination 
might cause in our final results and conclusions, the sample of 1533 stars with complete 
kinematic data has been put to several cleansings using the following sources:  the
observing notes taken at the telescope, the lists and catalogues of Carney (2003) and of
Carney et al.~(1994), and also the catalogues of data contained in the SIMBAD data base
(principally Dommanget \& Nys 1994).

The final data base cleaned of all binaries, suspected binaries, stars with contaminated
data as seen at the telescope, and probable variable stars contains 1223 high-velocity and
metal-poor stars.  This sample is seen in Figs.~4, 5, 6, and 7.  However, in some of the 
other analyses to follow, such as Figs.~8, 9, 10, 11, and 12, all of the 1533 stars are
plotted but with different symbols:  open symbols represent the most contaminated binaries,
the double-lined spectroscopic binaries (SB2 or dlsb); crossed-open symbols, all other
binaries, suspected binaries, or those noted at the telescope to be contaminated; 
asterisk-open symbols, the probable photometric variables; and solid symbols, stars which
suffer none of the above.

This binary-star symbolism is combined with that described above in Sect.~3.3 for the
$M_{\rm V}$s and distances, so that, for example, an SB2 star with its parallax taken
from Hipparcos with less than a 10\% error, is represented in the following figures with
an open circle.  A non-binary, non-variable star with its distance from the IDL-Hipparcos
empirical calibration, as a solid square, and so forth.  A more complete description
for the total symbolism of the following figures is given in the text to Fig.~8.

\section{[Fe/H],V(rot) diagram}

\subsection{Appearance and use}

In Fig.~5 a [Fe/H],V(rot) diagram is presented, similar to Fig.~5 of SPC but with more
stars, better kinematic data, better distances, and cleaned of the binary and variable
stars.  For this diagram V(rot) = V' + 220 km s$^{-1}$, and V' is the Galactic velocity in the
direction of Galactic rotation corrected for the solar motion, as described above in Sect.~4.2,
V' = V + 14.9 km s$^{-1}$, and $\Theta_{\rm 0} = 220$ km s$^{-1}$ is the assumed value of
the circular speed of the Milky Way at the solar position (Kerr \& Lynden-Bell 1986).  [Fe/H]
is the photometric metallicity as described in Sect.~3.2.  As in SPC, two main stellar
components are apparent in this Fig.~5, one centered at 
[Fe/H],V(rot) $\approx (-0.25$,+180 km s$^{-1}$), and the other, more dispersed, at
[Fe/H],V(rot) $\approx (-1.5,-25$ km s$^{-1}$).  The diagonal dashed line labeled ``0.0" is
that defined and used in Paper V and SPC to separate the halo stars (below and to the left)
from the ``high-velocity disk" stars (above and to the right).  The dot-dash line labeled
``LSR" indicates the level in V(rot) of the Local Standard of Rest (220 km s$^{-1}$), and the
other horizontal dot-dash line, V(rot) = 0 km s$^{-1}$.  Also shown is the direction of 
increasing X, the stellar population parameter defined in SPC as a linear combination of 
V(rot) and [Fe/H], and the loci with constant values of X = $-6.0$ and X = $-21.0$, the two 
diagonal dotted lines labeled with these values.  The criterion, $-21.0 \le$ X $\le -6.0$, is 
one of those to be used and discussed in this paper for defining the ``thick disk".  For
example, the two quadrilaterals of Fig.~5 define halo (X $\ge 0.00$) and ``thick disk"
($-21.0 \le$ X $\le -6.0$) samples, both with $-1.03 \le$ [Fe/H] $\le -0.63$, to be 
compared in $(b$--$y)_{\rm 0},M_{\rm V}$ diagrams for studying their relative ages.  In
general stars with more positive values of X are more halo-like, and stars with more
negative values, more disk-like.  As shown in SPC and in Fig.~6, the dashed line X = 0.00
provides a fairly good, but not totally clean, separation of the halo and disk stars; the
``halo" sample, below and to the right in Fig.~5, \emph {is} a nearly clean sample, largely
free of other stellar populations, while the ``high-velocity disk" stars, above and to the
right, are a mixture of thick-disk, old-thin-disk, and halo stars.
This X parameter shown in Fig.~5 has been used previously, such as in SPC and MS, to separate
several stellar populations and Galactic components and to study their relative-age 
differences.  

The discriminant X is useful for separating the stellar populations, but probability
algorithms given in the literature by Bensby et al.~2003 and by Venn et al.~2004 may
be superior.  In these, the stars' (U,V,W) velocity information is compared to the Galactic
Gaussian velocity-ellipsoid components, with assumptions concerning the velocity dispersions
and asymmetric drifts of the thin disk, thick disk, and halo, in order to estimate for each 
star its probability of membership in each of these populations.  Metallicity information, 
[Fe/H] or [m/H], is not used by these alternate methods.  The effectiveness of our X 
criterion for separating out thick-disk stars has been discussed and shown in Karata\c{s} et
al.~(2005).

%                                                One column figure
%-----------------------------------------------------------[Fe/H]-V(rot)
   \begin{figure}
   \centering
   \includegraphics[width=12cm]{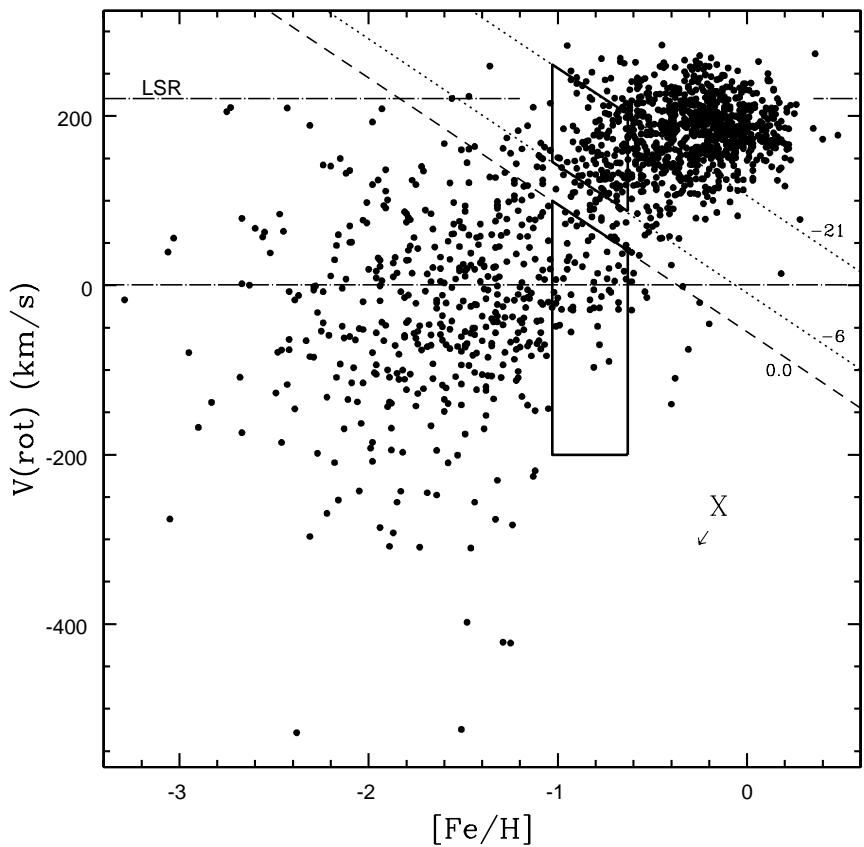}
      \caption{The [Fe/H],V(rot) diagram for 1223 stars with reliable
      kinematic parameters, cleaned of binary and variable stars.
      V(rot) = V' + 220 km s$^{-1}$, and the [Fe/H] values are from the
      calibration of Paper II.  The dashed diagonal line
      is that suggested in Paper V to separate halo from ``high-velocity disk''
      stars, and the direction of increasing ``X'', the stellar-population
      parameter defined in SPC, is shown.  The diagonal dotted lines
      correspond to X = $-6.0$ and $-21.0$, the limits used in this paper
      to define the ``thick disk''.  The horizontal dash-dot lines show
      V(rot) = 0 and +220 km s$^{-1}$, the latter corresponding to the
      Local Standard of Rest.  The two quadrilaterals show examples of
      samples (in this case for ``halo'' and ``thick-disk'' groups with
      $-1.03 \leq$ [Fe/H] $\leq -0.63$) to be compared to isochrones in
      the $(b$--$y)_{\rm 0}$,$M_{\rm V}$ diagram, such as in Fig.~12 below.
      }
         \label{FigFeHVrot}
   \end{figure}
%
%______________________________________________________________

\subsection{Histogram against X}

In Fig.~6 a histogram against X is presented, similar to Fig.~8b of SPC, with a bin
size of three units in X.  That is, the stars are counted in diagonal slices parallel
to the dashed line of Fig.~5 with intervals of three in X.  Again, a Gaussian fit to
the histogram has been attempted with three components, corresponding to the halo,
thick disk, and old thin disk.  Fig.~6 shows one of our better fits with the
normalization, mean value, and width of (N,$\langle$X$\rangle$,$\sigma_{\rm X}$) =
(26,23.0,19.0) for the halo component, (60,$-18.0$,8.0) for the thick disk, and
(122,$-29.0$,3.8) for the old thin disk\footnote{ The ratios of these normalizations do not reflect the local population densities of the halo, thick disk, and old thin disk,
but have been clearly biased by the largely kinematic selection criteria of our sample;
this is also true for the normalizations of Fig.~3.  Previous studies have generally
shown a local space density for the halo of 0.1--0.5\%, and for the thick disk of 2--11\%,
relative to the thin disk (for example, Gilmore 1984; Robin \& Cr\'ez\'e 1986; Sandage 1987; 
Yamagata \& Yoshii 1992; Reid \& Majewski 1993; Kerber et al.~2001; Larsen \& Humphreys 2003;
Karaali et al.~2004).}.  Clearly this Gaussian fit to the histogram
is not as good as that seen in Fig.~3 for the [Fe/H] histogram, nor as good as the
Gaussian fit to the X histogram presented in Fig.~8b of SPC.  The intermediate 
``thick-disk" Gaussian does not fit well the data, and there seems to be indications 
for two ``thick-disk" components, one as a shoulder to the halo distribution with 
X $\approx -8$, and the other as a shoulder to the old thin disk with X $\approx -20$.  
The sharp drop in the histogram at X = $-15$ is quite robust, showing up also in the 
histogram for the total sample of 1533 stars and in the various cleansings of the 
binary and variable stars.

%                                                One column figure
%-----------------------------------------------------------Xhisto
   \begin{figure}
   \centering
   \includegraphics[width=12cm]{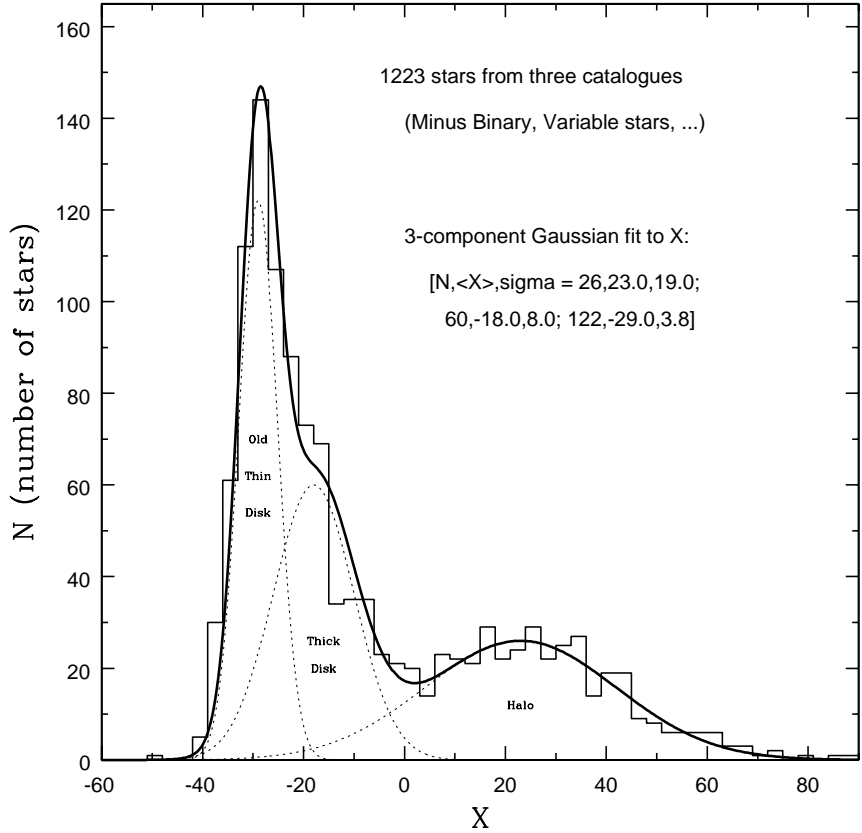}
      \caption{Histogram against X, the stellar-population parameter
      defined in Paper VI and shown in the previous figure.  Our total,
      cleaned sample of 1223 stars is included in this histogram.  A
      three-component Gaussian fit has been attempted taking mean
      values and dispersions from the literature for the halo,
      thick disk, and old thin disk.  With the improved kinematics
      and photometric distances of this paper, the Gaussian fit to the
      intermediate ``thick-disk'' X-values, $-21 \protect\la$ X $\protect\la -6$, is
      not nearly as good as it was in Fig.~8b of SPC; there is now a 
      serious discrepancy, a sharp drop in the counts, at X $\protect\approx -15$.}
         \label{FigXhisto}
   \end{figure}
%
%______________________________________________________________

\subsection{Two groupings in the thick disk?}

In Fig.~7 contour plots of the [Fe/H],V(rot) diagram are given, in the upper panel
for the entire figure, and in the lower panel concentrating more on the
``high-velocity-disk" stars at the more positive values of [Fe/H] and V(rot), i.e. more
negative values of X, as shown by the labels on the diagonal dotted lines.  In the
upper panel a kernel with a size of (0.40dex, 60 km s$^{-1}$) and a step size of (0.20dex,
30 km s$^{-1}$) has been used to sum over the [Fe/H],V(rot) diagram, and 100 levels of
contour plotted; in the lower panel a kernel with a size of (0.20dex, 30 km s$^{-1}$),
step size of (0.10dex, 15 km s$^{-1}$), and 40 contour levels.  In the upper panel, one 
can appreciate well again the two large, overall components mentioned above for Fig.~5, 
the halo population centered at [Fe/H],V(rot) $\approx (-1.3$dex,$-40$ km s$^{-1}$), and
a disk component at [Fe/H],V(rot) $\approx (-0.2$dex,+185 km s$^{-1}$).  It appears that
the median halo undergoes retrograde rotation in the Galaxy, assuming $\Theta_{\rm 0}$ =
220 km s$^{-1}$ for the circular speed of the Milky Way at the solar circle; if a more
modern value for this circular speed is used, such as the value of $\Theta_{\rm 0} =
234 \pm 13$ km s$^{-1}$ from Fukugita \& Peebles (2004), the median halo still remains
retrograde at about the two-sigma level.

%                                                One column figure
%-----------------------------------------------------------VrotFeHcontour
   \begin{figure}
   \centering
   \includegraphics[width=12cm]{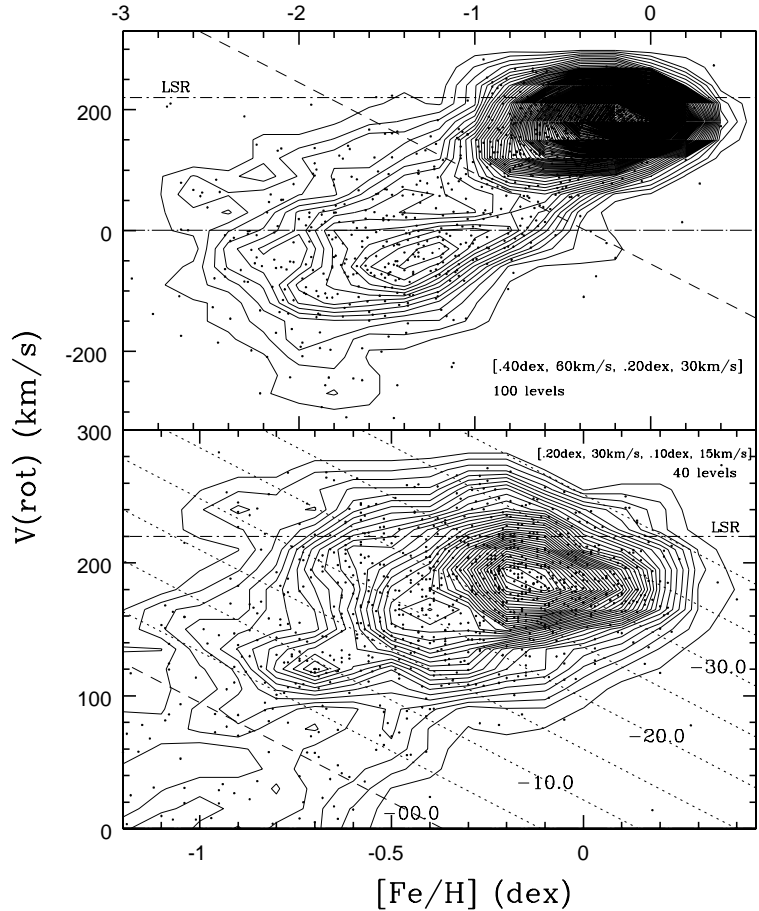}
      \caption{[Fe/H],V(rot) contours produced by passing
      a kernel over the full extent of the 1223 high-velocity and
      metal-poor stars, which are also plotted as small points.
      In the upper panel, the kernel size was (0.40 dex, 60 km s$^{-1}$)
      with a step size of (0.20 dex, 30 km s$^{-1}$), and in the lower
      a kernel of (0.20 dex, 30 km s$^{-1}$) with a step size of
      (0.10 dex, 15 km s$^{-1}$).  In the upper panel 100 equal-spaced
      levels have been shown over nearly the full extent of the halo
      and ``high-velocity disk'' stars, with only the sparsest regions
      not included by a contour.  The lower panel shows an expanded
      view of the ``high-velocity disk'' region with 40 equal-spaced
      levels.  The horizontal short-dash--dot line shows the V(rot) = 220
      km s$^{-1}$ value of the Local Standard of Rest (LSR); the
      horizontal long-dash--dot line, V(rot) = 0 km s$^{-1}$; the
      diagonal dashed line, the loci of X = 0.00, suggested in
      Paper V to separate halo from ``high-velocity disk'' stars; and
      the diagonal dotted lines the loci with constant X values of
      $-5.0, -10.0, -15.0, -20.0$, ..., as labeled.
      }
         \label{FigVrotFeHcontour}
   \end{figure}
%
%______________________________________________________________

The lower panel of Fig.~7 shows evidence of additional structure, i.e. stellar components, 
for the ``high-velocity-disk" stars, X $< 0.00$.  The main peak at [Fe/H],V(rot) $\approx
(-0.2$dex,+185 km s$^{-1}$) corresponds to the old thin disk stars, but two other
components are noted with [Fe/H],V(rot) $\approx (-0.4$dex,+160 km s$^{-1}$)
and [Fe/H],V(rot) $\approx (-0.7$dex,+120 km s$^{-1}$).  The division between these two
latter groups gives the sharp drop at X = $-15$ in the X histogram of Fig.~6.  In
other words, what was called the ``thick disk" in previous papers may have two
components.  To check that these two components really are ``thick", the dispersions
in the vertical Galactic velocity W' have been examined.  For the more disk-like
component, its limits have been set at $-23.0 \le$ X $\le -16.0$ and
$-0.60 \le$ [Fe/H] $\le -0.20$; its vertical dispersion in the W' velocity is
45.2 km s$^{-1}$ from 146 stars drawn from the total sample of 1533 stars, and 45.8
km s$^{-1}$ for 125 stars drawn from the cleaned sample of 1223.  For the more halo-like
component, its limits have been taken at $-14.0 \le$ X $\le -6.0$ and
$-0.90 \le$ [Fe/H] $\le -0.50$; its vertical dispersion is 61.3 km s$^{-1}$ from 83 stars
drawn from the total sample, and 62.0 km s$^{-1}$ for 64 stars from the cleaned sample.
These dispersions for the more halo-like, thick-disk component
are probably contaminated by halo stars, as can be seen in the Gaussian fit of
Fig.~6, but these dispersions do clearly show that these two thick-disk components
really are ``thick" and not just physical extensions of the old thin disk; SPC and MS 
argued that the maximum possible vertical dispersion for the old thin disk lies in the 
range 30--35 km s$^{-1}$, based on works such as that of Wielen (1977) or Freeman (1991).

\section{$(b$--$y)_{\rm 0}$,$M_{\rm V}$ diagrams and stellar ages}

\subsection{The thick disk}

In Fig.~8 are given $(b$--$y)_{\rm 0}$,$M_{\rm V}$ diagrams for the two proposed
components of the thick disk, one with $-23.0 \le$ X $\le -16.0$ and
$-0.60 \le$ [Fe/H] $\le -0.20$ (the more disk-like), and the other with
$-14.0 \le$ X $\le -6.0$ and $-0.90 \le$ [Fe/H] $\le -0.50$ (the more halo-like).
The $M_{\rm V}$ values have been derived as described above in Sect.~3.3, from
Hipparcos, from our empirical photometric calibration, and from the older
calibration of Paper V\footnote{Papers by Furhmann (1998), Prochaska et al.~(2000),
Feltzing et al.~(2003), Reddy et al.~(2003), Bensby et al.~(2003, 2005), and Brewer \&
Carney (2004) show clearly that the thin disk and thick disk obey different
[$\alpha$/Fe] versus [Fe/H] relations.  In as much as the $(b$--$y)$, $m_{\rm 1}$, and
$c_{\rm 1}$ photometric indices do not track [$\alpha$/Fe], this will produce
increased scatter in our photometric M$_{\rm v}$ calibration.  However, this difference
will not produce systematic effects in our Figs.~8, 12, and 13, since we are comparing
only thick-disk, or thick-disk plus halo, stars in these figures.}.  
The $(b$--$y)_{\rm 0}$ values have been dereddened
as described in Sect.~3.1.  The isochrones are those of Bergbusch \& VandenBerg
(2001) as transformed to $uvby$ photometry by Clem et al.~(2004).  In the upper
panel the isochrones corresponding to [Fe/H] = $-0.40$ and [$\alpha$/Fe] = +0.00
have been plotted, and in the lower, [Fe/H] = $-0.70$ and [$\alpha$/Fe] = +0.30.
These [$\alpha$/Fe] values seem most appropriate considering the mean relation
between [$\alpha$/Fe] and [Fe/H] seen in field stars for [Fe/H] $> -1.0$ (M\'arquez
2005; Bressan, A.~2004; Barbuy et al.~2003; Clementini et al.~1999; McWilliam 1997).
The error bars correspond to a 10\% distance error and
to a typical single observation in $(b$--$y)$; many of the stars will have errors
smaller by a factor of the square root of two or three.  The symbols of Fig.~8
combine the notations for $M_{\rm V}$ and for the binary/variable stars, as 
described at the end of Sects.~3.3 and 5, and in the text to the figure; the
symbols representing the most reliable data are the solid circles (non-binary,
non-variable stars with $M_{\rm V}$ from Hipparcos, and a less than 10\% parallax
error), and the solid squares (non-binary, non-variable stars with $M_{\rm V}$
from our empirical Hipparcos photometric calibration).

%                                                One column figure
%-----------------------------------------------------------DiskGroupIsos
   \begin{figure}
   \centering
   \includegraphics[width=12cm]{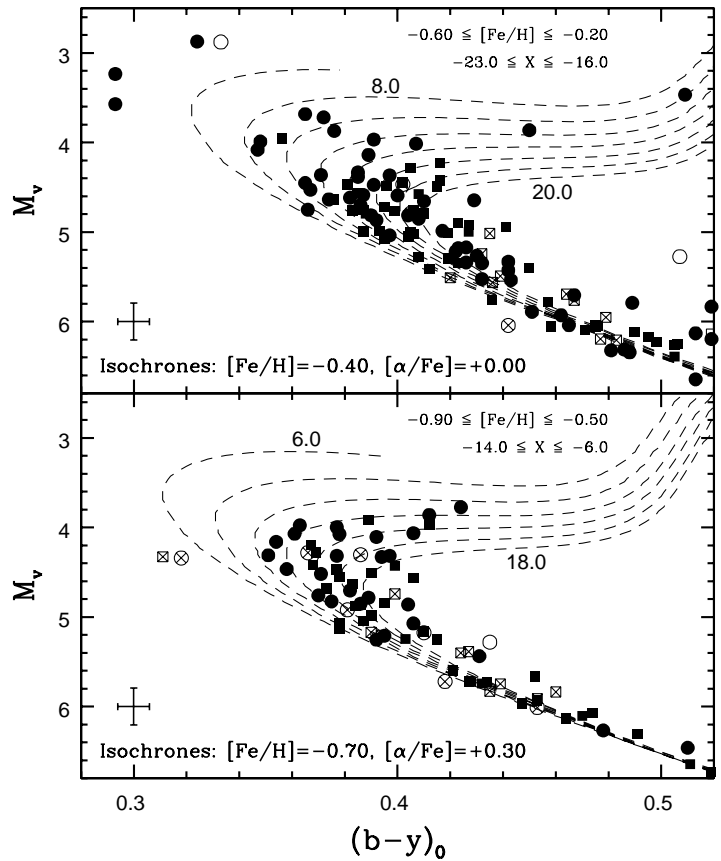}
      \caption{$(b$--$y)_{\rm 0}$,$M_{\rm V}$ diagrams for two
      of the disk groups identified in the previous figure, one
      with [Fe/H] $\approx -0.40$ and $-23 \protect\la$ X $\protect\la -16$,
      and the other with [Fe/H] $\approx -0.70$ and $-14 \protect\la$ X $\protect\la -6$.
      Isochrones from Bergbusch \& VandenBerg (2001; Clem et al.~2004) 
      are plotted in steps of
      2 Gyr from 6 Gyr to 18 or 20 Gyr, and error bars corresponding
      to a single observation are shown; most of the stars plotted
      have three or more photometric observations.  For the symbols:
      all circles correspond to $M_{\rm V}$ values direct from the
      Hipparcos parallaxes with errors less than 10\%; all squares
      to $M_{\rm V}$ from our photometric calibration; and all
      triangles to other quality $M_{\rm V}$ values, such as from
      Hipparcos parallaxes with errors larger than 10\% or from the 
      calibration of Paper V.  All solid symbols show non-binary stars; 
      open symbols, SB2 binaries; crossed-open, other binary stars; 
      and the asterisk-open, the photometrically variable stars.
      }
         \label{FigDiskGroupIsos}
   \end{figure}
%
%______________________________________________________________

In the upper panel of Fig.~8 (the more disk-like component) a wide range of ages
is present, as might be expect from Fig.~7.  This group lies where there is 
significant overlap between the thick disk and the old thin disk, with probably
a few halo stars thrown in also.  A few stars are seen with ages less than 6 Gyr,
and some with quite large ages, $\ga 13$ Gyr, but the most obvious sequence is 
defined for a group of 8--10 stars with ages $\approx$ 9--11 Gyr.
In the lower panel of Fig.~8 the scatter is much less, and the more obvious
sequence corresponds to an age $\approx$ 11--14 Gyr.  Below in Fig.~13, it will
be seen that the halo population of the Galaxy has a mean age of $13.0 \pm 0.2$
Gyr.  So, the lower panel of Fig.~8 shows that part of the thick disk has an
age very similar to that of the halo.  Such a conclusion has already been
obtained by MS using field stars, and by Venn et al.~(2005), who use globular 
clusters with measured space velocities to assign memberships in either the halo 
or thick disk, finding similar ages for the two cluster groups and, hence, for the
two populations.  Also, Fig.~8 shows that the two components of the thick disk of 
Figs.~6--7 have ages differing by about 2.5 Gyr.

\subsection{Younger metal-poor stars?}

In Fig.9 a [Fe/H],$(b$--$y)_{\rm 0}$ diagram is plotted, similar to Fig.~11 of
Schuster et al.~(1996) but with more stars and better, more recent isochrones,
those of Bergbusch \& VandenBerg (2001; Clem et al.~2004)
In the upper panel only the stars and error bars are plotted
without the isochrones.  The error bars correspond to typical errors in [Fe/H]
and in $(b$--$y)_{\rm 0}$ at the more metal-rich and more metal-poor limits of
this figure.  The open circles mark the halo stars (X $> 0.00$) from the SN,
SPC, and present catalogues of high-velocity and metal-poor stars.  The open 
triangles show the VMP stars from Paper X and from Schuster et al.~(1996); only
VMP stars classified TO, MS, SG, or BS-TO (the turn-off, main sequence, and
subgiant stars) have been plotted. An obvious lower ridge-line is seen, whose
$(b$--$y)_{\rm 0}$ varies with [Fe/H], as one would expect.  In the lower panel
one can see that this ridge-line corresponds to $13 \pm 1$ Gyr.  The isochrones
given are those for [$\alpha$/Fe] = +0.30, which is appropriate for the more
metal-poor stars, [Fe/H] $\la -0.8$ (M\'arquez 2005; Bressan, A.~2004; 
Barbuy et al.~2003; Clementini et al.~1999; McWilliam 1997; Nissen \& Schuster 
1997; Carney 1996; Carney et al.~1997)).

%                                                One column figure
%----------------------------------------------------------by0-[Fe/H]
   \begin{figure}
   \centering
   \includegraphics[width=12cm]{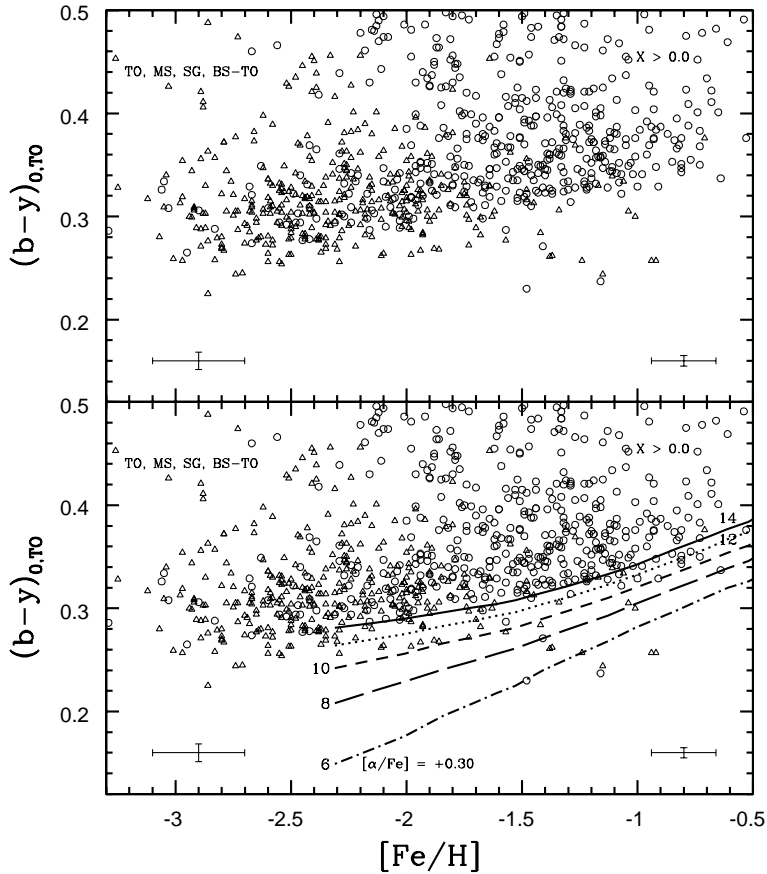}
      \caption{The [Fe/H],$(b$--$y)_{\rm 0}$ diagram for the high-velocity
      and metal-poor stars of this paper (open circles) plus the
      very-metal-poor stars from Schuster et al.~(2004) (open triangles).
      The [Fe/H] values come from the calibration of Paper II for the
      former stars, and
      from calibrations of the HK survey, as described in Schuster et
      al.~(2004), for the latter.  The error bars show the calibration
      errors for [Fe/H], as given in the above sources, and the error of
      a single observation for $(b$--$y)_{\rm 0}$; most of the
      high-velocity and metal-poor stars have been observed three or more
      times, and the very-metal-poor stars 1-3 times.  For the former
      group only halo stars (X $> 0.0$) have been included, and for the
      latter group only the turn-off, main-sequence, and subgiant stars
      (TO, MS, SG, and TO-BS) have been plotted, as classified in
      Schuster et al.~(2004).  In the lower panel the turn-off loci
      from the isochrones of Bergbusch \& VandenBerg (2001; 
      Clem et al.~2004) are over-plotted for
      the ages of 6, 8, 10, 12, and 14 Gyr, with [$\alpha$/Fe] = +0.30.
      }
         \label{Figby0FeH}
   \end{figure}
%
%______________________________________________________________

A number of stars bluer than this ridge-line are noted, especially over the
metallicity range, $-1.5 \la$ [Fe/H] $\la -0.9$, where there are stars with
ages reaching $\approx$ 8 Gyr younger than the ridge-line.  These are probably 
stars analogous to the blue, metal-poor (BMP) stars of Preston et 
al.~(1994) with [Fe/H] $< -1.0$ and $0\fm15 < (B-V)_{\rm 0} < 0\fm35$, bluer 
than the globular cluster turnoffs.  At even lower metallicities there are 
stars with ages $\approx 3$ Gyr younger than the ridge-line, and this is very 
similar to differences discussed for Figs.~10 and 11 of Paper X, where the 
VMP stars were compared to the turn-off of the globular cluster M92.

These BMP stars have been discussed in the literature, for example
Preston et al.~(1994) and Unavane et al.~(1996), as being the result of
accretion events by the Galaxy of metal-poor, intermediate-age dwarf
spheroidal satellites.  But they have also been interpreted, for example
by Preston \& Sneden (2000) and by Carney et al.~(2005), as containing a
large fraction, ``at least half'', of blue stragglers.

\subsection{Stellar groups in the $(b$--$y)_{\rm 0}$,$M_{\rm V}$ diagram}

In Figs.~10--12 are shown the $(b$--$y)_{\rm 0}$,$M_{\rm V}$ diagrams for
three stellar-population, metallicity-range groups.  Figures 10 and 11 show
``halo" groups (X $> 0.00$) according to the stellar population parameter
defined in SPC and shown above in Fig.~5; Fig.~12 presents a ``thick-disk" group
according to $-21.0 \le$ X $\le -6.0$.  This latter thick-disk range assumes
that the thick disk is a single stellar population, as shown for example in 
Figs.~1b and 8b of SPC and in Fig.~3 of this paper, but contradicted
by Figs.~6 and 7.  The cuts at X = $-21.0$ and X = $-6.0$ serve to reduce the
contamination by other stellar populations, such as the halo and old-thin-disk, 
as shown in Fig.~6 of this paper and Fig.~8b of SPC, while retaining an 
adequate sample size.  This ``thick-disk" definition gives 305 stars  with a
vertical velocity dispersion of $\sigma_{\rm W'} = 52.6$ km s$^{-1}$ from our
total sample of 1533 stars, and 246 stars with $\sigma_{\rm W'} = 53.1$ km s$^{-1}$
from the cleaned sample of 1223 stars.  As mentioned above, these vertical velocity
dispersions are probably larger than the actual true dispersion of the thick
disk, due to contamination by the halo, which is obvious in Fig.~6.

%                                                One column figure
%-----------------------------------------------------Mv,by0-halo-1.71
   \begin{figure}
   \centering
   \includegraphics[width=12cm]{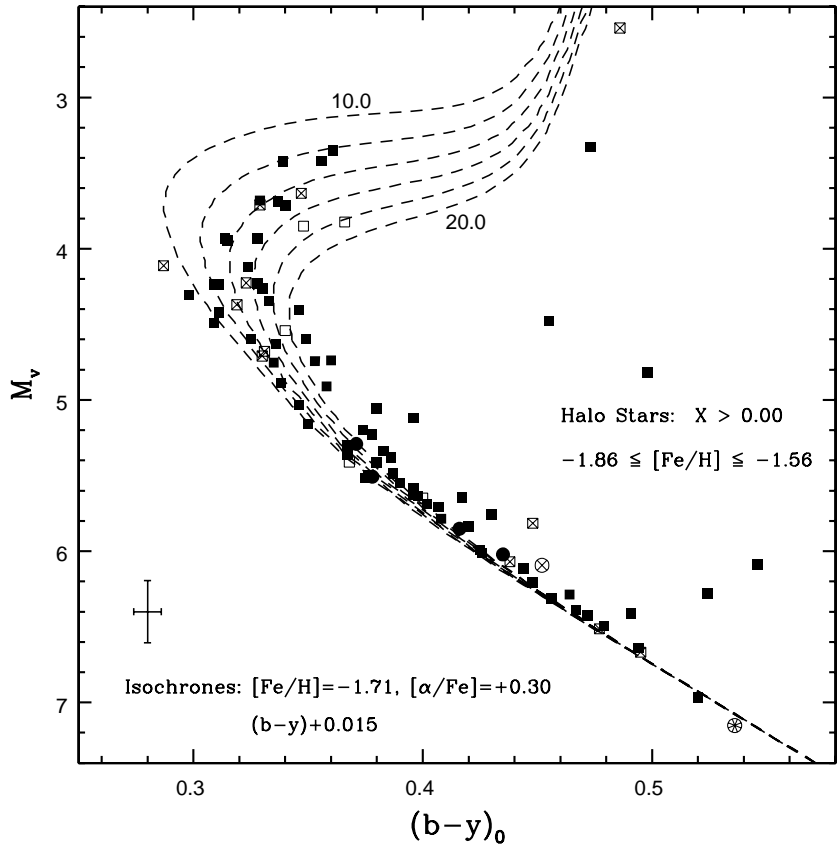}
      \caption{The $(b$--$y)_{\rm 0}$,$M_{\rm V}$ diagram for
      halo stars (X $>$ 0.00) with metallicities of
      [Fe/H] $\approx -1.71$.  The symbols are the same, the
      isochrones from the same sources, and the error bars defined
      in the same way as for Fig.~8.  The stars plotted (minus the
      binary and variable stars) have a mean value for [Fe/H] of
      $-1.71$, exactly the same value as for the isochrones.  The
      isochrones have been shifted by +0.015 in $(b$--$y)_{\rm 0}$
      to provide a better fit for the redder, main-sequence, 
      little-evolved stars.
      }
         \label{FigMvby0halom171}
   \end{figure}
%
%______________________________________________________________

%                                                One column figure
%-----------------------------------------------------Mv,by0-halo-1.31
   \begin{figure}
   \centering
   \includegraphics[width=12cm]{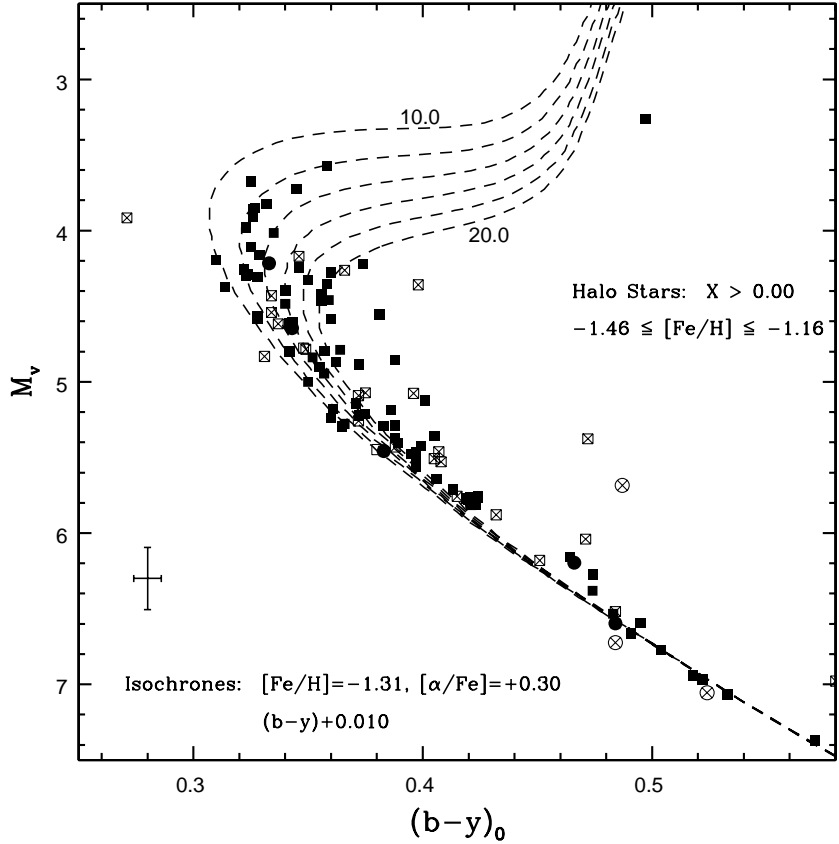}
      \caption{The $(b$--$y)_{\rm 0}$,$M_{\rm V}$ diagram for
      halo stars (X $>$ 0.00) with metallicities of
      [Fe/H] $\approx -1.31$. The symbols are the same, the
      isochrones from the same sources, and the error bars defined
      in the same way as for Fig.~8.  The stars plotted (minus the
      binary and variable stars) have a mean value for [Fe/H] of
      $-1.31$, exactly the same value as for the isochrones.  Here
      the isochrones have been shifted by +0.010 in $(b$--$y)_{\rm 0}$
      to provide a better fit for the redder, main-sequence,
      little-evolved stars.
      }
         \label{FigMvby0halom131}
   \end{figure}
%
%______________________________________________________________

%                                                One column figure
%-----------------------------------------Mv,by0-thickdisk-0.83
   \begin{figure}
   \centering
   \includegraphics[width=12cm]{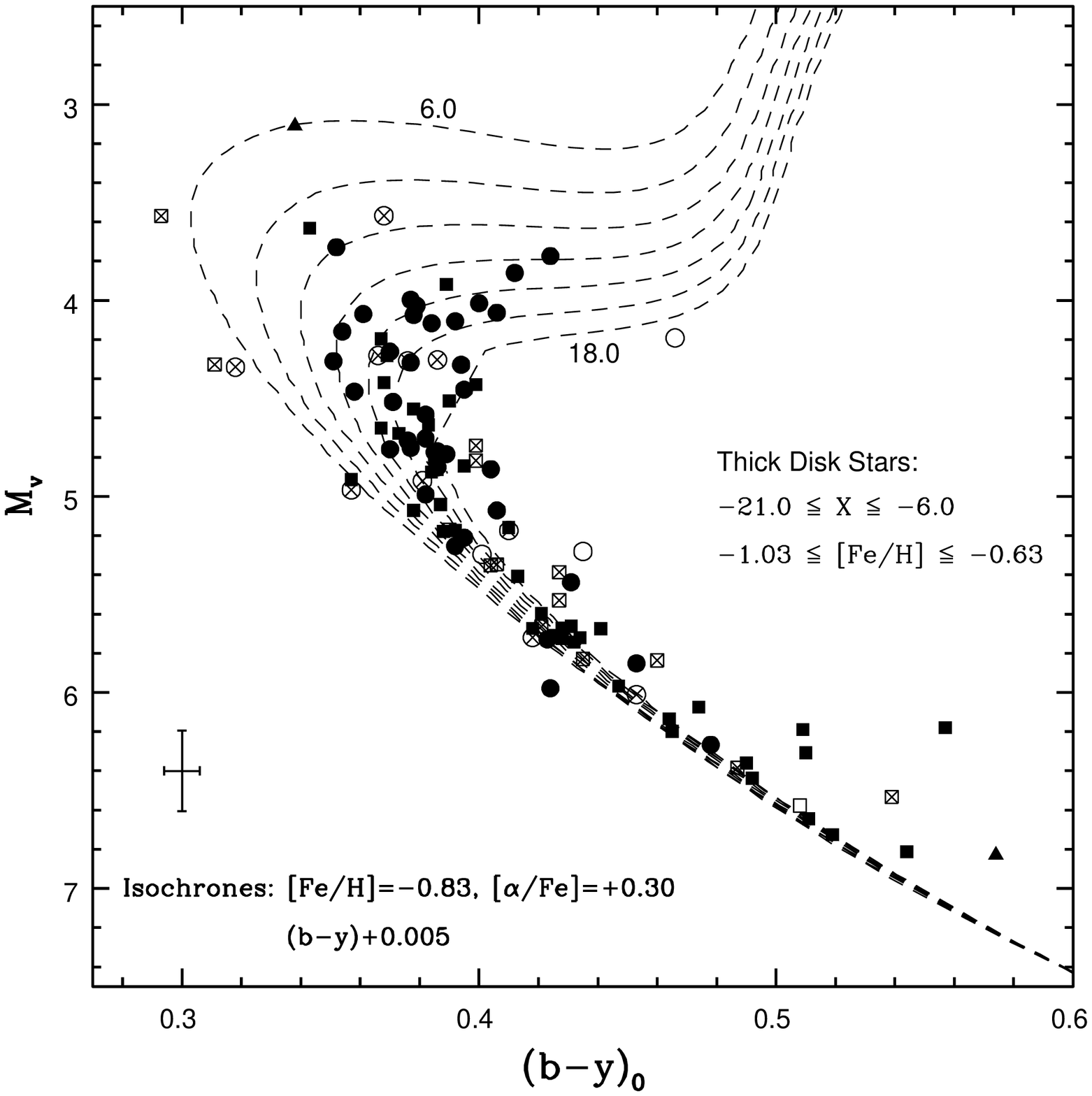}
      \caption{The $(b$--$y)_{\rm 0}$,$M_{\rm V}$ diagram for
      thick disk stars ($-21.0 \leq$ X $\leq -6.0$) with metallicities
      of [Fe/H] $\approx -0.83$.  The symbols are the same, the
      isochrones from the same sources, and the error bars defined
      in the same way as for Fig.~8.  The stars plotted (minus the
      binary and variable stars) have a mean value for [Fe/H] of
      $-0.76$, slightly larger than that of the isochrones.  The
      isochrones have been shifted by +0.005 in $(b$--$y)_{\rm 0}$
      to provide a better fit for the redder, main-sequence, 
      little-evolved stars.
      }
         \label{FigMvby0thickdisk}
   \end{figure}
%
%______________________________________________________________

In Fig.~10 halo stars with $-1.86 \le$ [Fe/H] $\le -1.56$ are compared to
isochrones with [Fe/H] = $-1.71$; these are the same isochrones as used above
in Figs.~8 and 9, those of Bergbusch \& VandenBerg (2001; Clem et al.~2004).
In Fig.~11 halo stars with
$-1.46 \le$ [Fe/H] $\le -1.16$ are compared to isochrones with [Fe/H] = $-1.31$,
and in Fig.~12, thick-disk stars with $-1.03 \le$ [Fe/H] $\le -0.63$ are compared
to isochrones with [Fe/H] = $-0.83$.  In all these cases [$\alpha$/Fe] = $+0.30$
has been used, which is appropriate, on the average, for stars with [Fe/H] $\la -0.8$
(M\'arquez 2005; Bressan, A.~2004; Barbuy et al.~2003; Clementini et al.~1999;
McWilliam 1997; Nissen \& Schuster 1997; Carney 1996; Carney et al.~1997).  For 
metallicities in the range $-0.83 \le$ [Fe/H] $\le -0.30$, the [$\alpha$/Fe] values
used to determine the ages of the stellar groups have been interpolated between 
+0.30 and 0.00, as a function of [Fe/H], according to a mean relation for field 
stars developed by M\'arquez (2005) using spectroscopic values from several recent
sources, as mentioned above.  Also noted in Figs.~10--12, the isochrones have all 
been corrected slightly in $(b$--$y)_{\rm 0}$ so that the lower theoretical main
sequence of the isochrones, $(b$--$y)_{\rm 0} \ga +0.40$, matches more closely the 
lower observed main sequence.  These small corrections, $0\fm005$--$0\fm015$ are 
merely fine adjustments to the very carefully derived color-$T_{\rm eff}$ relations 
of Clem et al.~(2004).

In Fig.~10 a clear sequence of non-binary stars is seen with a mean age of
$13.4 \pm 1.1$ Gyr, from the six most evolved halo stars; in Fig.~11, $12.0 \pm 1.0$
Gyr, from the eight most evolved, non-binary, halo stars; and in Fig.~12,
$14.1 \pm 1.2$, from the ten most evolved, non-binary, thick-disk stars.  In Fig.~12
contamination from another stellar population (the old-thin-disk) is seen with 
stars as young as 6.0 Gyr, but these have not been included in our average age nor 
in the statistics.  The errors given above are standard deviations for a single 
determination (single star) within each sequence, assuming that each sequence is in fact
isochronic, which may not strictly be the case.  These estimated errors are smaller 
when the evolving sequence is well defined with more evolving stars and little 
contamination from other stellar populations, and larger when the evolving sequence 
is not well defined with few evolving stars and more contamination, as seen below in
Fig.~13.

\section{Age comparisons}

\subsection{Age versus [Fe/H] diagram}

In Fig.~13 the [Fe/H],age diagram is plotted for 12 halo groups (X $\ge 0.0$)
over the metallicity range $-2.31 \le$ [Fe/H] $\le -0.83$, and for 7 "thick
disk" groups ($-21.0 \le$ X $\le -6.0$) over the metallicity range $-1.01 \le$
[Fe/H] $\le -0.30$.  These metallicity ranges are defined in part by the range
for the isochrones of Bergbusch \& VandenBerg (2001; Clem et al.~2004)
($-2.31 \le$ [Fe/H] $\le -0.30$) and in part by the intervals in the stellar 
population parameter X.  For example, in Fig.~5 halo stars (X $\ge 0.0$) can 
be seen to metallicities as high as [Fe/H] $\approx -0.2$, but there are far 
too few of these for [Fe/H] $\ga -0.8$ to define any sort of age sequence.  
The same for ``thick disk'' stars for metallicities less than 
[Fe/H] $\approx -1.0$.

%                                                One column figure
%-------------------------------------------------------Age-[Fe/H]
   \begin{figure}
   \centering
   \includegraphics[width=12cm]{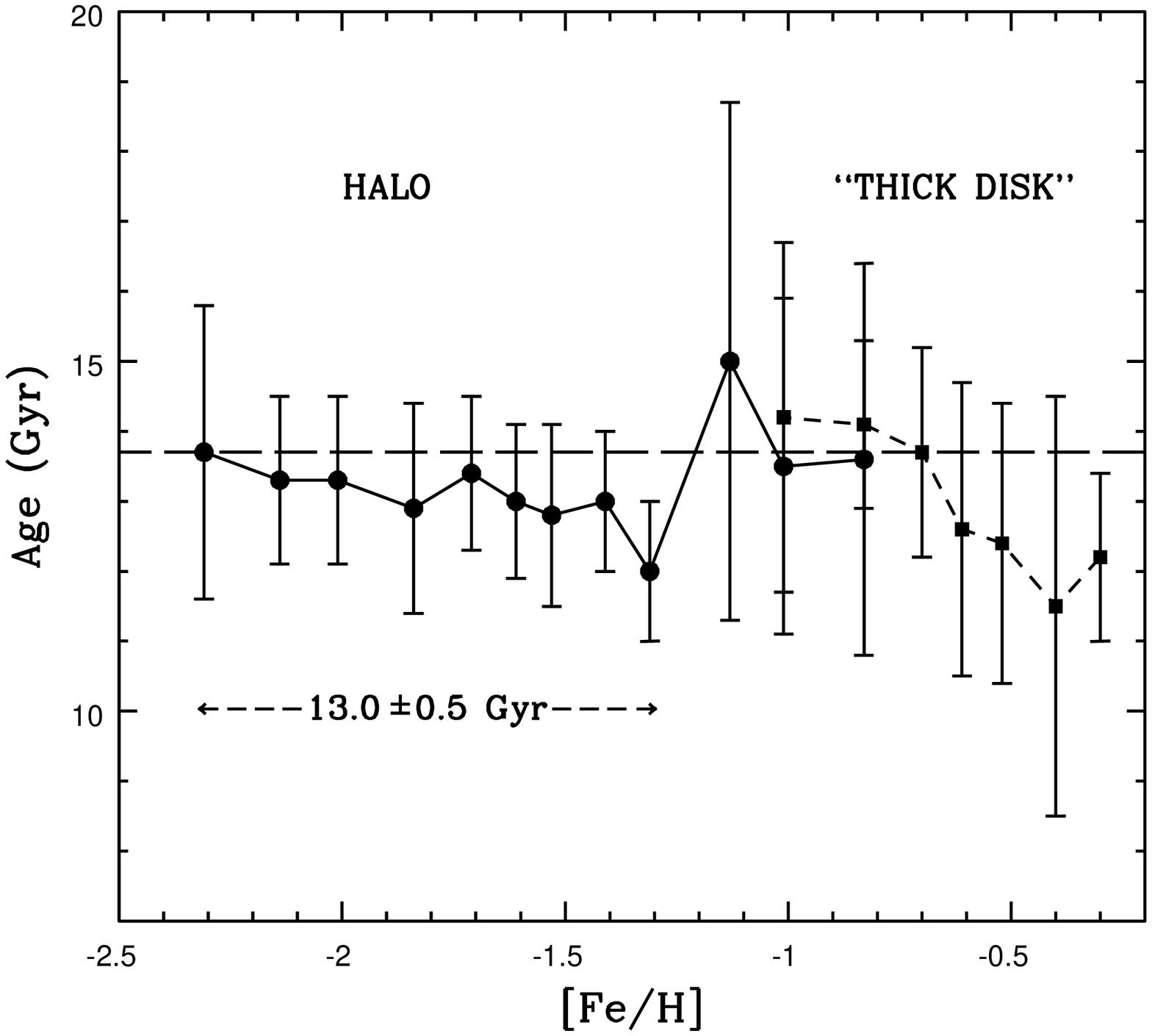}
      \caption{The [Fe/H],age diagram for halo and ``thick disk''
      stars.  The ages have been derived from $(b$--$y)_{\rm 0}$,$M_{\rm V}$
      diagrams, as in Figs.~10--12.  The ages and error bars have been
      estimated from the 4--10 most evolved stars in each diagram, and
      the smaller error bars result from those cases where a clear
      evolving sequence is noted; outliers probably from other stellar
      populations, as seen in Fig.~12, have been excluded.  These error
      bars represent the estimated error for a single determination
      (single star), and so should be divided by factors of 2--3 for a
      mean error.  The solid circles and solid line show the results for
      the halo stars, as defined in the [Fe/H],V(rot) diagram of Fig.~5
      (X $\ge 0.0$), and the solid squares and dashed line the results for
      the ``thick disk'' stars from the same [Fe/H],V(rot) diagram
      ($-21.0 \le$ X $\le -6.0$).  The horizontal dashed line corresponds
      to 13.7 Gyr from Bennett et al.~(2003), the WMAP results for the age
      of the Universe.  The nine most metal-poor ``halo" points give an
      average of 13.0 Gyr with a scatter of $\pm 0.5$ Gyr and a mean error
      of about $\pm 0.2$ Gyr.
      }
         \label{FigAgeFeH}
   \end{figure}
%
%______________________________________________________________
%

Most of the error bars of Fig.~13 are in the range, 
$1.0 \la \sigma_{\rm age} \la 2.0$ for a single determination (single star)
except where there are too few evolving stars to define a good sequence or
where there is contamination by other stellar populations.  For example,
at [Fe/H] $= -2.31$ the halo sample should not be suffering any contamination
by disk stars, but there are few ($\approx 4$) evolving stars to define a good
age sequence, and these show more scatter perhaps due to increasing influence
by the detailed chemical abundances, such as [N/Fe] and [C/Fe], upon the $uvby$ 
photometry (see Paper X).  Over the range $-1.13 \le$ [Fe/H] $\le -0.83$ the halo
samples are suffering decreasing numbers of stars, few of which are evolved,
and perhaps some small contamination by the thick disk; the error bars are
larger, 1--4 Gyr.  Over the range $-0.61 \le$ [Fe/H] $\le -0.40$ the ``thick
disk'' samples show larger error bars (2--3 Gyr) probably due to several 
reasons:  there are fewer than eight evolving stars for each sample; probable small
contaminations by both the old-thin-disk and halo, as seen in Fig.~6, so that 
the evolving sequences are not well defined; and also the differing ages 
between the two ``thick disk'' components discussed above in Sects.~6.3 and 7.1.

In Fig.~13 the halo stars show a slight decrease in mean age over the range
$-2.31 \le$ [Fe/H] $\le -1.31$ with an average age of 13.0 Gyr and a single-value
standard deviation of $\pm 0.5$; this reduces to approximately $\pm 0.2$ for the
mean error of the nine points.  The age-metallicity relation which is seen for the
halo points in Fig.~13 is not significant considering the single-value error bars,
and only marginally significant for the mean-value error bars.  Over the range
$-1.13 \le$ [Fe/H] $\le -0.83$, this age-metallicity relation breaks down completely,
and the error bars have grown considerably for the reasons mentioned in the
previous paragraph:  few evolved stars and perhaps some small contamination by
the thick disk.  

The horizontal dashed line of Fig.~13 corresponds to 13.7 Gyr from Bennett et 
al.~(2003), the WMAP results for the age of the Universe.  The results of Fig.~13
would then suggest that the first stars of the Galaxy formed $0.7 \pm 0.3$ Gyr
after the creation of the Universe, combining our mean error with that of the
WMAP value.  For this combined mean error, we have assumed that the systematic 
errors of the isochrones are negligible, based on the continuing high-quality work 
of Bergbusch \& VandenBerg (2001) for stellar models and isochrones, the very 
carefully derived empirical color-temperature relations of Clem et al.~(2004) for 
$uvby$ photometry, and the smallness of the color corrections that we have applied,
as seen in Figs.~10-12.  Time and continued use of these isochrones and 
color-temperature relations will test this assumption.  Also, the supposition that
[$\alpha$/Fe] = +0.30 for the halo stars may introduce a systematic difference; if
in fact [$\alpha$/Fe] = +0.35 is more realistic for the halo stars over $-2.3 \la$ 
[Fe/H] $\la -1.3$, as suggested in some studies (M\'arquez 2005), the mean halo age 
of Fig.~13 would drop by about another 0.3 Gyr.

The ``thick-disk'' points of Fig.~13 show a very strong age-metallicity correlation,
but again this is not highly significant considering the single-value error bars.
And, this correlation can mostly be explained as due to varying contributions of 
the two ``thick-disk'' components discussed above in Sects.~6.3 and 7.1, one with
([Fe/H],V(rot),X) $\approx (-0.7$ dex,+120 km s$^{-1}$,$-9.0$) and the other with
([Fe/H],V(rot),X) $\approx (-0.4$ dex,+160 km s$^{-1}$,$-21.0$).  For the points of
Fig.~13 the ages have been determined using metallicity intervals of $\pm 0.15$ or
$\pm 0.20$ dex in [Fe/H], the wider range being used to include more stars when
needed in the $(b$--$y)_{\rm 0}$,$M_{\rm V}$ diagram.  So, for the ``thick-disk'' 
cut which we are using in Fig.~13, $-21.0 \le$ [Fe/H] $\le -6.0$, at [Fe/H] = 
$-1.01$, $-0.83$, and $-0.70$ the first of the two above ``thick-disk'' components 
dominates, and this has the larger age of 11--14 Gyrs, as shown in Fig.~8b and as 
discussed in Sect.~7.1.  At [Fe/H] = $-0.40$ and $-0.30$ the second ``thick-disk'' 
components dominates, and this has the smaller age of 9--11 Gyrs as shown in 
Fig.~8a.  For metallicities in between there are varying contributions of these 
two components to the mean ages plotted in Fig.~13.

\subsection{Episodes during the Galactic formation and evolution?}

The previous figures and discussions point to several evidences for different
episodes or physical processes during the formation and evolution of the Galaxy.
The most obvious of these, which has been discussed amply in the literature, is
the dichotomy seen in Figs.~5 and 7a, between the Galaxy's halo and disk.  For
example, in Fig.~5 the oval concentration of stars centered at
[Fe/H],V(rot) $\approx (-0.25$,+180 km s$^{-1}$) with ranges of approximately
$\pm 70$ km s$^{-1}$ in V(rot) and approximately $\pm 0.5$ dex in [Fe/H],
corresponds to the ``high-velocity disk'' as discussed in Paper V and in SPC.
This component of Fig.~5 is made up of mostly disk stars, old-thin and thick,
with some small contamination by the halo, as indicated in Fig.~6.  The second
obvious component of Fig.~5 is that of the halo, more dispersed, centered at
[Fe/H],V(rot) $\approx (-1.5,-25$ km s$^{-1}$) with ranges of approximately
$\pm 250$ km s$^{-1}$ in V(rot) and approximately $\pm 1.35$ dex in [Fe/H].
As seen in Fig.~6 and in Fig.~8b of SPC, this stellar component is mostly
uncontaminated by the other stellar populations.

But in Figs.~6, 7, and 9 there are other indications of episodes during the
Galactic evolution:  within the ``high-velocity disk'' from Figs.~6 and 7, and
within the halo from Fig.~9.  The sharp drop in the number of stars at
X $\approx -15.0$, ([Fe/H],V(rot)) $\approx (-0.5$--$-0.6$ dex,140--150 km s$^{-1}$),
seen in the histogram of Fig.~6, indicates the separation of two events or physical
processes at values of metallicity and velocity were the ``thick disk'' has
normally been studied (see SPC).  The contours of Fig.~7b show two probable
components between the peaks of the old-thin disk and of the halo; these are
separated at X $\approx -15.0$.  The chemodynamic galaxy formation code of 
Brook et al.~(2004) proposes that the thick disk was formed during an epoch of
 ``...multiple mergers of gas-rich building blocks...'', that such a scenario 
 fits best the observed kinematic, chemical, and age characteristics of the
Galaxy's thick disk.  Perhaps what is seen in Figs.~7 and 8 is evidence for 
two such accretion events, one about 12.5 Gyr ago and another about 10 Gyr.

In Fig.~9, the $(b$--$y)_{\rm 0}$,[Fe/H] diagram shows evidence for at least 
two episodes, or intervals, in the evolution of the Galactic halo.  Over the 
metallicity range $-1.5 \la$ [Fe/H] $\la -0.9$ there are numerous stars which 
are younger by as much as 8 Gyr than the ridge-line defined by the large 
majority of halo stars.  For [Fe/H] $\la -1.5$ these younger stars are much 
less obvious, but there is a number of stars up to 3 Gyr younger than this
ridge-line.  Preston et al.~(1994) have argued that their blue, metal-poor 
stars, those with [Fe/H] $< -1.0$ and $0\fm15 < (B-V)_{\rm 0} < 0\fm35$, bluer 
than the globular cluster turnoffs, are evidence for the accretion by the 
Galaxy of stars from other dwarf galaxies.  (Preston \& Sneden (2000)
and Carney et al.~(2005) would argue that many of these BMP stars are in fact
blue stragglers.)  Also, in Paper X arguments have been made that the younger very
metal-poor stars ([Fe/H] $\la -1.5$) present evidence for the hierarchical
star-formation/mass-infall of very metal-poor material to the Galaxy and/or
for the accretion of material from other (dwarf) galaxies with different
formation and chemical-enrichment histories.  However, the spectroscopic
study of Venn et al.~(2004) ``...rules out continuous merging of low-mass
galaxies similar to these dSph satellites during the formation of the
Galaxy.''  The detailed chemical abundances of the local dwarf galaxies are
too different from those of the stellar components of the Galaxy.  Only for
[Fe/H] $\le -1.8$ might such a scenario work, since for this range the
chemistries of the dSphs are in fair agreement with the Galactic halo stars.
It would interesting to observe spectroscopically whether the bluer, metal-poor
stars of Fig.~9, those with $-1.5 \la$ [Fe/H] $\la -0.9$, have detailed
abundances more like the local dSphs or more like the other stellar components
of the Galaxy.  Also, it would be interesting to observe these BMP stars
of Fig.~9 for evidence of binary stars with low-eccentricity orbits, 
higher-than-average rotational velocities, and lithium deficiencies, all
indications for probable blue stragglers (Preston \& Sneden 2000; Carney et 
al.~2005).

In any case Fig.~9 indicates some sort of main formation event
for the Galaxy 13--14 Gyr ago, such as a collapse, as shown by the ridge-line,
and then formation of the more metal-poor stars, [Fe/H] $\la -1.5$, for as long
as 3 Gyr afterwards, and, perhaps, of less metal-poor stars, 
$-1.5 \la$ [Fe/H] $\la -0.9$, for as long as 8 Gyr afterwards.

\subsection{Galactic and cosmological implications}

\subsubsection{The thick disk}

As mentioned above and as seen in Figs.~6, 7, and 8, evidence for two episodes
(or Galactic components) between the old thin disk and halo can be detected in 
our data base of high-velocity and metal-poor stars.  These stellar thick-disk 
components have ([Fe/H], V(rot), X, Age, $\sigma_{\rm W'}) \approx (-0.7$ dex, 
120 km s$^{-1}$, $-9.0$, 12.5 Gyr, 62.0 km s$^{-1}$) and $\approx (-0.4$ dex, 
160 km s$^{-1}$, $-21.0$, 10.0 Gyr, 45.8 km s$^{-1}$), with some contamination 
to these values by the other stellar populations, especially the halo to the 
former group as can be seen in Fig.~6.  Parker et al.~(2003, 2004) and Gilmore 
et al.~(2002) also find evidence for two components within the thick disk.  
Parker et al. use both star-count and kinematic evidence to argue for the 
existence of an ``asymmetric thick disk'' with significant differences between 
the quadrants I and IV of the Galaxy.  Quadrant I has a significant (20-25\%) 
excess of stars over the counts in quadrant IV, and a rotation lag of 80--90 
km s$^{-1}$ compared to a lag of only 20 km s$^{-1}$ for quadrant IV.  They
mention three possibilities to explain this ``asymmetric thick disk'':  the fossil
remnant of a galaxy merger, a triaxial thick disk, or an interaction between the
thick disk and the Galactic bar; they prefer the latter explanation but hedge
their final conclusions mentioning the recently discovered debris stream in Canis
Major (Martin et al.~2004).  Vel\'azquez \& White (1999) have proposed the 
formation of asymmetric disks as the result of galactic accretion processes.

Gilmore et al.~(2002) also find evidence for two probable components within the
thick disk by studying stars 0.5--5.0 kpc from the Galactic plane.  Surprisingly
they find a V(rot) a few kpc above the plane of only about 100 km s$^{-1}$ 
compared to the expected 180 km s$^{-1}$, i.e. an excess lag of about 
80 km s$^{-1}$, and conclude that this is probably evidence for a merger event
with the disk of the Milky Way some 10--12 Gyr ago, that their sample is
dominated by the remnants of a disrupted satellite galaxy.  These values for
the rotational lag and age agree quite well with our values given above for the
first thick-disk component mentioned above.  According to the arguments of Gilmore 
et al., our component with [Fe/H] $\approx -0.70$ would represent the remnants of
a shredded satellite from an early merger, while the component with 
[Fe/H] $\approx -0.40$ the more ``classical'' thick disk or perhaps a later merger
that provided the main mass of the local thick disk.

The discovery of the probable satellite remnant in Canis Major by Martin et 
al.~(2004; also Bellazzini et al.~2004, Martin et al.~2005, Bellazzini et al.~2005,
Dinescu et al. 2005) provides very serious observational confirmation and explanation for 
the above results of Gilmore et al. and perhaps also those of Parker et 
al\footnote{The reality of the Canis Majoris dwarf galaxy is still in
doubt; Momany et al.~2004 propose that Martin et al. are really looking at the 
southern Galactic warp, and Carraro et al.~2005 argue that this Canis Majoris
``dwarf galaxy'' and its Blue Plume are really sub-structures in the
Norma-Cygnus and Perseus spiral arms and their inter-arm region.}.  This
debris stream orbits very close to the Galactic plane, with a pericenter close to
the solar distance, R$_\odot$, and with an orbital eccentricity and vertical
scale height similar to that of the thick disk.  Martin et al. conclude that the
Galactic thick disk is continually growing, even up to the present epoch, and that
dwarf galaxies on nearly co-planar orbits, such as the precursor of the Canis
Majoris remnant, form the main building blocks of the Galaxy's thick disk.  Also, 
Yoachim \& Dalcanton (2005) have observed counterrotating thick disks in external 
galaxies and claim that this too strongly supports the idea that thick disks form 
from the direct accretion of infalling satellite galaxies.  Yong et al.~(2005)
discuss the unusual elemental abundance ratios of the open clusters in the outer
Galactic disk, and propose that the formation of these outer clusters may have
been triggered by a series of merger events in the outer disk, such as the Canis
Majoris event.  These results point to 
the conclusion that one, or perhaps both, of the thick-disk components documented 
above in our Figs~6--8 are due to satellite remnants, debris streams, such as that 
observed in Canis Major.  The disrupting dwarf galaxy discovered in Sagittarius by 
Ibata et al.~(1994) also provides such possibilities for structure within the 
Galaxy's halo and thick disk.

The Galactic models given and documented by Kroupa (2002), Abadi et al.~(2003),
and Brook et al.~(2004) also provide support for probable structure within the
thick disk, such as that described above.  The work of Kroupa predicts the thickening
of the Galactic disk through ``clustered star formation'' induced by the interaction
of the Galactic disk with passing satellite galaxies.  He proposes that star clusters 
are the ``building blocks'' of the Galaxy, that massive clusters add kinematically 
hot components to the Galactic field stellar populations, and that for ages $>3$ Gyr 
ago the Galaxy suffered a period with an ICMF (initial cluster mass function) extending 
to very massive clusters.  This scenario is very useful in light of the 
chemical-difference results of Venn et al.~(2004) comparing the detailed abundances 
of the local dwarf galaxies with those of the thick disk in that direct infall of the 
passing satellite is not needed to heat the disk.  Under such a scheme, the two 
thick-disk components seen in Fig.~7 might have been caused by two major interactions 
of the Galactic disk with passing satellite galaxies.

The models of Abadi et al. and Brook et al. both make use of hierarchical clustering
(merging) within a $\Lambda$CDM universe and obtain similar results.  They do not
specifically model the Milky Way but a very similar late-type galaxy, and produce
results very interesting for explaining the thick disk and also the behavior shown in
our Figs.~6--8.  The model of Abadi et al. produces starbursts (their figure 7)
triggered by merger and accretion events with major episodes occurring at 8.5, 10.0,
11.5, and 13.0 Gyr ago.  They predict that about 15\% of the thin disk and more than
60\% of the thick disk are made up of tidal debris from satellite galaxies; for ages
older than $\approx 10$ Gyr, about 90\% of the thick disk shares this origin.  For 
this model the tidal disruption of satellite galaxies is seen as the main process of 
assembly of their simulated thick disk, within distinct episodes.  The ``chemodynamical''
galaxy formation code of Brook et al. produces the thick disk at ages greater than
$\approx 8$ Gyr during a very chaotic period during the end of the peak SFR.  The 
thick-disk stars are formed from ``gas-rich building blocks''  accreted to the galaxy 
during such a period that is the ``natural consequence'' of the early, violent 
hierarchical clustering of the $\Lambda$CDM universe.  They predict that thick
disks should be prevalent for disk galaxies.  The structure seen in our Figs.~6--8
would suggest two rather significant ``gas-rich building blocks'' or accretion episodes
for the Milky Way thick disk.

However, it is more difficult to accommodate these latter two models to the results of 
Venn et al.~(2004), who compared the detailed chemical signatures of the Galactic 
stellar populations with those of the local dwarf galaxies.  They conclude that the 
thick disk is not comprised of remnants of low-mass dSph galaxies nor of remnants of
higher-mass dwarf galaxies such as the LMC or the one in Sagittarius, ``because of
differences in chemistry.''  However, the accretion of gas-rich components, such as
in the model of Brook et al., rather than actual stars, may circumvent this discrepancy.

\subsubsection{The bluer, metal-poor stars}

In our Fig.~9 a very distinct difference is seen between the blue limits over the
range $-1.5 \la$ [Fe/H] $\la -0.9$ as compared to the blue limits over 
[Fe/H] $\la -1.5$.  The majority of halo stars (X $> 0.0$) and very-metal-poor stars
form a ridge-line corresponding to 13--14 Gyr, while for the former [Fe/H] interval
the bluest stars are $\approx 8$ Gyr younger but only about 2--3 Gyr younger over the
latter interval (see also Figs.~10--11 of Paper X).  These stars are very analogous 
to the BMP stars of Preston et al.~(1994), which have 
[Fe/H] $\la -1.0$ and fall blueward of the most metal-poor globular cluster turnoffs, 
0.15 $\la$ (B--V)$_{\rm 0} \la$ 0.35.  Preston et al. find an isotropic velocity 
dispersion for the BMPs, $\sigma_{r\phi\theta} \approx 90$ km s$^{-1}$ and conclude 
that these BMP stars probably were accreted to the Galaxy from dwarf spheroidal 
satellites during the last $\approx 10$ Gyr.  But why the dichotomy at [Fe/H] $\approx
-1.5$?

Figure 2 of Unavane et al.~(1996), (B--V)$_{\rm 0}$ versus [Fe/H], is very similar to
our Fig.~9.  Their figure does not distinguish between halo and thick-disk stars but
plots ``metal-poor'' stars from the proper-motion-selected sample of Carney et al.~(1994).
Again, a ridge-line is seen corresponding to an age of 15--16 Gyr (according to the
revised Yale Isochrones) with bluer stars extending $\approx 8$ Gyr to younger ages over
the interval $-1.5 \la$ [Fe/H] $\la -1.0$ and only 2--3 Gyr younger for [Fe/H] $\la -1.5$,
very analogous to our Fig.~9, except for a 2 Gyr offset in the absolute ages.
They estimate that this younger-star population makes up about 5\% of the stars in the
most metal-poor range, [Fe/H] $< -1.95$, about 8\% in the intermediate range, $-1.95 <$
[Fe/H] $< -1.5$, but about 34\% over the range $-1.5 <$ [Fe/H] $< -1.0$.  They also
compare these bluer stars to the turnoff colors seen in local dSph galaxies and try to
estimate the possible number of mergers by Carina-like galaxies ($\approx 60$) or by 
Fornax-like dwarfs ($\approx 6$) to produce these BMP stars.

Seen in terms of a hierarchical-clustering scheme and, for example, Fig.~7 of Abadi et
al.~2003, Fig.~9 shows that the initial clustering was rather abrupt, as shown by the
ridge-line, tapering off rather quickly for 2-3 Gyr afterwards.  Then later there was
an episode (or episodes) such as the accretion of a dwarf galaxy (galaxies) that
brought in material with [Fe/H] $\approx -1.2 \pm0.3$ dex, forming or bringing stars 
as much as 8 Gyr younger than the typical halo star.  One might conclude that the 
youngest of these stars formed during the accretion event ("starburst") itself, about 
5--6 Gyr ago.  It would interesting to study the detailed chemical abundances of the BMP
stars seen in Fig.~9, those with $(b$--$y) \la 0.30$ and $-1.5 \la$ [Fe/H] $\la -0.9$:
BD$+25\degr1981$, BD$-12\degr2669$, G202--065, BPS BS 15621--0070, BPS BS 16026--0073,
BPS BS 16081--0038, BPS BS 17444--0059, BPS BS 17581--0075, BPS BS 17581--0077,
BPS CS 22887--0048, BPS CS 30311--0068, and BPS CS 30493--0001, and interpret 
these in the light of the conclusions made by Venn et al.~(2004).  Do these stars have
chemical abundances more like those of the Galactic stellar populations, or more like
those of the local dwarf galaxies?  Was the formation of these stars induced by the close
passage of a satellite galaxy (the clustered star formation of Kroupa 2002), or were they
actually brought into the Galaxy by the merger/accretion of the satellite (Abadi et 
al.~2003)?  Alternatively, do they show the characteristics of blue stragglers?  The
first three stars listed above (BD$+25\degr1981$, BD$-12\degr2669$, G202--065) have been
studied as probable metal-poor field blue stragglers by Carney et al.~2005.  (The 
coordinates and characteristics of the above BPS stars are given in the tables of Paper X,
also available at the CDS).

\subsubsection{The oldest stars}

Figure 13, a plot of mean stellar ages versus [Fe/H], derived from 
$(b$--$y)_{\rm 0}$,$M_{\rm V}$ diagrams such as those in Figs.~10--12, 
shows that in general our ages are in agreement
with the WMAP results for the age of the Universe (Bennett et al.~2003), $13.7 \pm 0.2$ 
Gyr.  The halo groupings agree very well, with an average $13.0 \pm 0.2$ Gyr (mean error).
Over the range $-1.2 \la$ [Fe/H] $\la -0.8$ some of our ages are larger than the WMAP 
result, but this is the interval where the stellar populations (especially the halo and 
thick disk) are most mixed and difficult to separate, where the BMP-star contribution to
the halo groupings is most important (Sect.~8.3.2), where the stellar sequences in the 
$(b$--$y)_{\rm 0}$,$M_{\rm V}$ diagrams are least well defined, and where the estimated
errors of the groupings are the largest, $\pm2$--4 Gyr.  None of the largest ages in 
Fig.~13 are discrepant with the WMAP result at the $2\sigma$ level (mean errors); in 
fact, the most incongruent is only about $1\sigma$ greater than the WMAP age.

From the nine most metal-poor halo groupings of Fig.~13, we estimate a time interval
of $0.7 \pm 0.3$ Gyr from the beginning of the Universe to the formation of the Galactic
halo.  This compares quite favorably with the estimate of 0.8--1.0 Gyr that Krause (2003,
2004) and Krauss \& Chaboyer (2003) estimate using globular cluster ages.  The WMAP 
results (Bennett et al.~2003) also indicate that reionization occurred at an age of
0.1--0.4 Gyr ($180^{+220}_{-80}$ Myr), leaving 0.6--0.3 Gyr for the formation of the
Galactic Population II, which is entirely feasible considering that there is considerable
agreement that the zero-metallicity/very-low-metallicity Population III had a high-mass
heavy IMF (Bromm et al.~1999, 2002; Abel et al.~2000; Nakamura \& Umemura 2001; Hernandez
\& Ferrara 2001; Bromm 2004; Omukai et al.~2005); these studies all point to the first
generation stars having mostly stars in the range 10--100M$_{\odot}$, or even higher.  The
zero-metallicity stellar models of Marigo et al.~2001 indicate that such massive, metal-free
stars, in the range 10--100M$_{\odot}$, have nuclear lifetimes of 3$\times 10^6$ to
2$\times 10^7$ yrs.
So, there is more than enough time for the zero-metallicity/very-low-metallicity Pop. III
stars to evolve, enrich the interstellar medium, and set the stage for the Pop. II,
more-metal-rich stars to follow.  These results also corroborate Krause (2003) and provide
another ``hint'' for a time interval between reionization and the formation of larger-scale
structures, such as the Galactic halo and globular clusters.

Also seen in Fig.~13 are mean ages for thick disk stars, those selected using the criterion
$-21.0 \le$ X $\le -6.0$ from the [Fe/H],V(rot) diagram.  Some of these thick-disk groups
are seen to be as old as the halo stars, which agrees with the hierarchical clustering
models of Abadi et al.~(2003) and Brook et al.~2004.  For example in the case of Abadi et
al., the stars in the outermost, presumably most metal-poor regions of the accreting
satellites contribute primarily to the Galaxy's
halo and thick disk, since they can be more easily stripped, while those in the core are 
more resistant to disruption, have more time to have their orbits circularized by dynamical
friction, and so end up in the old thin disk.  About 50\% of the spheroid and 60\% or more
of the thick disk share their origin in the same merger/accretion events; for the older
thick disk (ages $\ga 10$ Gyr) about 90\% of the stars were brought to the galaxy in this way.
So, many of the halo and thick-disk stars participated in the same merger/accretion events
and should have similar ages.

The thick-disk component of Fig.~13 also shows a very clear age-metallicity relation with
the more metal-poor thick-disk stars ([Fe/H] $\approx -1.0$) being $\approx 3$ Gyr older
than the higher-metallicity ones ([Fe/H] $\approx -0.3$).  This agrees, at least qualitatively,
with the age-metallicity relation found for the Galactic thick disk by Bensby et al.~(2004).
Their Figs.~5 and 8 show a somewhat larger range in ages, 4--5 Gyr, but they are also working
over a somewhat greater range in [Fe/H], $-0.90$ to 0.00.  Their largest ages for the
thick disk, 12--15 Gyr, agree well with our largest as seen in Fig.~13, but their smaller
ages, 6--10 Gyr, are less than ours, due mostly to the fact that they are able to track
the thick disk to [Fe/H]=0.0, while we work only to the limit of the available isochrones,
[Fe/H]=$-$0.30.  At [Fe/H]=$-$0.30 Bensby et al.~(2004) obtain ages of 8--11.5 Gyr for the
thick disk, slightly lower than in our Fig.~13.  Also, their method for separating out
the thick-disk stars is more rigorous, and strictly kinematic, making use of the
velocity dispersions and asymmetric drifts for the three stellar populations (thin disk, 
thick disk, and halo), while our method
uses only the X criterion, a function of [Fe/H] and V(rot) as described above.  And, much of
our age-metallicity relation is produced by the different mean characteristics of the two
thick-disk components discussed above in Sect.~8.3.1; one has (age,[Fe/H]) $\approx$ (12.5
Gyr, $-0.7$ dex) and the other $\approx (10.0, -0.4)$.  In any case, we would agree mostly
with at least two of the main conclusions of Bensby et al.:  that the star formation of the
thick disk was ongoing for several Gyr, and that the thick disk was the result of merger and/or
interaction events with satellite galaxies.  The Fig.~7 of Abadi et al.~2003 shows the 
thick-disk phase of the galaxy formation ongoing for 5--6 Gyr.

In many of the $(b$--$y)_{\rm 0}$,$M_{\rm V}$ diagrams, such as Figs.~8, 10, 11, and 12,
stars are plotted which appear to have ages greater than 14--16 Gyr, sometimes much greater.
The mean ages plotted in Fig.~13 generally represent well-formed sequences at the younger
extreme of these $(b$--$y)_{\rm 0}$,$M_{\rm V}$ diagrams, but frequently ``older'' stars are
present.  For example, in Fig.~11 a sequence of eight stars is seen which follows the
20-Gyr isochrone, as well as a number of stars which appear even older, and not all of
these have been identified as binaries.  If these stars really are this old would impose
serious cosmological implications!  But a number of other possibilities have to be considered:
(a) unidentified binaries; certainly not all binaries have been recognized through photometric
and radial-velocity studies; (b) larger than mean observational errors; the error bars which
are shown represent $\pm 1 \sigma$ error, and approximately 5\% of the stars will have errors
in $M_{\rm V}$ and $(b$--$y)_{\rm 0}$ larger than $2 \sigma$; (c) unusual chemical
compositions and their effects upon the photometric indices; as discussed in Paper X, the
$c_{\rm 0}$ index may be shifted by anomalous carbon and nitrogen abundances via the NH, CH,
and CN molecular bands, and these effects will increase with decreasing temperature.  These 
possibilities suggest a number of future studies, such as detailed, careful, and extended
spectroscopic studies of the radial velocities and detailed abundances of those stars which
appear older than $\approx$14--16 Gyr.  To confirm or deny these large ages would have
important cosmological connotations.

\section{Conclusions}

   \begin{enumerate}
      \item The catalogue of $uvby$--$\beta$ values presented in Table 1
         for 442 high-velocity and metal-poor stars is closely on the same
	 photometric systems as the previous catalogues of SN and SPC (see
	 Figs.~1 and 2).
	 
      \item The photometric [Fe/H] histogram for our total sample of 1223
         high-velocity and metal-poor stars can be fit very well using only
	 three Gaussian components with 
	 ($\langle$[Fe/H]$\rangle$,$\sigma_{\rm [Fe/H]}$) = ($-1.40$,0.60) 
	 for the halo Gaussian, ($-0.55$,0.18) for the thick disk, and 
	 ($-0.16$,0.14) for the old thin disk (see Fig.~3).
	 
      \item Our empirical, photometric calibration for $M_{\rm V}$ based on
         512 Hipparcos stars with parallax errors less than 10\% and cleaned 
	 of binaries, has a scatter of $\pm 0\fm206$ and shows no significant 
	 systematic problems over the
	 metallicity range $-2.39 \le$ [Fe/H] $\le 0.00$, as shown in Fig.~4.
	 
      \item Our X histogram is not fit well by a three-component Gaussian
         as previously in SPC, where X is our stellar population parameter,
	 a linear combination of [Fe/H] and V(rot).  A sharp drop in the
	 number of stars at X = $-15.0$ falls near the values of ([Fe/H],
	 V(rot),X) where the thick disk had previously been studied in SPC, and
	 may show evidence for structure within the thick disk.  (See Fig.~6).
	 A contour plot of the [Fe/H],V(rot) diagram shows two probable 
	 components between the old-thin disk and the halo, one with ([Fe/H],
         V(rot), X, Age, $\sigma_{W'}$) $\approx$ ($-0.7$ dex, 120 km s$^{-1}$,
	 $-9.0$, 12.5 Gyr, 62.0 km s$^{-1}$), and the other with $\approx$ 
	 ($-0.4$, 160, $-20.0$, 10.0, 45.8).  (See Figs.~7 and 8).  This agrees
	 well with the results of Gilmore et al.~(2002) and with Parker et 
	 al.~(2003; 2004) concerning components within the thick disk.
	 
      \item The [Fe/H],$(b$--$y)_{\rm 0}$ diagram for halo stars (Fig.~9) shows
         different bluest limits between the metallicity ranges,
	 $-1.5 \la$ [Fe/H] $\la -0.9$ and [Fe/H] $\la -1.5$.  In the former 
	 interval the bluest stars extend to about 8 Gyr younger than the 
	 ridge-line defined by most of the halo stars, while in the second 
	 interval there are halo stars only about 2--3 Gyr younger than this 
	 ridge-line.  One could conclude that the initial hierarchical clustering
	 was rather abrupt, tapering off rather quickly for 2-3 Gyr afterwards,
	 and then about 5--6 Gyrs ago an accretion/merger event brought in material
	 with [Fe/H] $\approx -1.2 \pm0.3$ dex.  (The bluer stars of this
	 metallicity interval have also been interpreted as field blue stragglers.)

      \item The present results show considerable evidence for episodes during the 
         formation and evolution of the Galaxy and its stellar populations.  The 
	 [Fe/H],V(rot) diagram of Fig.~5 shows the very well-known dichotomy 
	 between the disk and halo, with the halo having an age of about 13 Gyr
	 (Fig.~13).  Figs. 6--8 show probable structure within the thick disk, 
	 with two components present having ages of about 10 and 12.5 Gyr.  The
	 [Fe/H],$(b$--$y)_{\rm 0}$ diagram of Fig.~9 presents evidence for 
	 significant differences as a function of [Fe/H] for the formation of the 
	 youngest halo stars, with a probable event having occurred $\approx5.5$ Gyr 
	 ago.  All this supports the $\Lambda$CDM hierarchical-clustering scheme, 
	 such as the model of Abadi et al.~(2003), which produces starbursts 
	 triggered by  merger/accretion events at 8.5, 10.0, 11.5, and 13.0 Gyr for 
	 a galaxy model similar to the Milky Way.
	  
      \item Our [Fe/H],age diagram (see Fig.~13) indicates that the more metal-poor
         stars analyzed here, over the range $-2.31 \la$ [Fe/H] $\la -1.31$,
	 have a mean age of 13.0 Gyr, only $0.7 \pm 0.3$ Gyr younger than the age
	 of the Universe given by the WMAP results of Bennett et al.~(2003).  This
	 result agrees well with those of Krause (2003, 2004) that the oldest
	 globular clusters formed at about 0.8--1.0 Gyr, and that larger-scale
	 structures, such as the Galactic halo and globular clusters, formed after
	 reionization.  An interval of $0.7 \pm 0.3$ Gyr is sufficient for the
	 formation and complete evolution of zero-metallicity/very-low-metallicity 
	 Population III stars, thought to have had a very high-mass heavy IMF.

      \end{enumerate}

\begin{acknowledgements}
      W.J.S. is very grateful to the DGAPA--PAPIIT (UNAM) (projects Nos. 
      IN101495 and IN111500) and to CONACyT (M\'exico) (projects Nos. 
      1219--E9203 and 27884E) for funding which permitted travel and also 
      the maintenance and upgrading of the $uvby$--$\beta$ photometer.  A.M.
      acknowledges financial support from FCT (Portugal) through grants
      BPD/20193/99 and SFRH/BPD/19105/2005; and CONACyT (Mexico; project
      I33940-E).  W.J.S. also thanks Jos\'e Guichard, who extended the 
      invitation to spend a sabbatical year at INAOE, where much of the 
      analysis for this publication has been done, I. Aretxaga, M. Plionis,
      E. Gazta\~naga, ... for an interesting and useful cosmology
      discussion group, and M. Reyes Mu\~nos, G. Hernandez Palacios, and
      A. L\'opez of the computing center of INAOE for much needed 
      help.  Don VandenBerg and James Clem made their 
      $uvby$ isochrones available prior to publication, and we greatly 
      appreciate it.  Bruce W. Carney provided A. M\'arquez and W. J. 
      Schuster with information concerning the binary-star contamination 
      of their samples prior to publication, and we sincerely thank him.
      We thank Xavier Hernandez, who helped with useful ideas, discussions,
      and references.  Many people at the SPM observatory have helped over 
      the years; we thank especially L. Gut\'{\i}errez, V. Garc\'{\i}a 
      (deceased), B. Hern\'andez, J.L. Ochoa, J.M. Murillo, F. Quiros, 
      E. Colorado, F. Murillo, J. Valdez, B. Garc\'{\i}a, B. Mart\'{\i}nez, 
      E. L\'opez, A. Cordova, M.E. Jim\'enez, and G. Puig.  This publication 
      has made use of the SIMBAD database, operated at CDS, Strasbourg, 
      France.  We would also like to thank Bruce Carney, the referee, for
      many ideas and references used to polish the final manuscript.
         
\end{acknowledgements}

\end{document}